\documentclass[pra, nofootinbib, reprint, amsmath, amssymb, aps, showkeys,
superscriptaddress]{revtex4-1}
\pdfoutput=1
\usepackage{graphicx}% Include figure files
\usepackage{dcolumn}% Align table columns on decimal point
\usepackage{bm}% bold math
\usepackage{subcaption}
\usepackage{color,soul}
\usepackage{textcomp}
\usepackage[section]{placeins}
\usepackage{booktabs, siunitx}
\usepackage{colortbl}
\usepackage{xcolor}
\usepackage{xfrac}
\usepackage{array}
\usepackage{amsmath}
\usepackage{amsthm}
\usepackage{amssymb}
\usepackage[section]{placeins}
\usepackage[bottom]{footmisc}
\usepackage{hyperref}
\usepackage[capitalise]{cleveref}

\soulregister\cite7
\soulregister\ref7
\soulregister\pageref7

\newcommand{\pbar}{\overline{p}}

\newcommand{\bh}{{\bf h}}

\newcommand{\bx}{{\bf x}}

\newcommand{\bB}{{\bf B}}
\newcommand{\bv}{{\bf V}}
\newcommand{\bP}{{P}}
\newcommand{\hc}{{r,\phi,\zeta}}

\newcommand{\ph}{\phi}
\newcommand{\z}{\zeta}

\newcommand{\nnorm}[1]{{\left\lVert#1\right\rVert}}

\newcommand{\diff}[2]{\dfrac{\partial #1}{\partial #2}}
\newcommand{\ddiff}[2]{\dfrac{\partial^2 #1}{\partial {#2}^2}}
\newcommand{\mdiff}[3]{\dfrac{\partial^2 #1}{\partial {#2} \partial {#3}}}

\newcommand{\MAT}[1]{\renewcommand*{\arraystretch}{1.25} \begin{bmatrix} #1 \end{bmatrix}}

\begin{document}

    \title{Modeling Magnetic Fields with Helical Solutions to Laplace's Equation}%

    \author{Brian Pollack}
    \affiliation{Department of Physics and Astronomy,
        Northwestern University, Evanston, Illinois, 60208, USA}
    \author{Ryan Pellico}
    \affiliation{Department of Mathematics,
        Trinity College, Hartford, Connecticut, 06106, USA}
    \author{Cole Kampa}
    \affiliation{Department of Physics and Astronomy,
        Northwestern University, Evanston, Illinois, 60208, USA}
    \author{Henry Glass}
    \affiliation{Fermi National Accelerator Laboratory,
        Batavia, Illinois, 60510, USA}
    \author{Michael Schmitt}
    \affiliation{Department of Physics and Astronomy,
        Northwestern University, Evanston, Illinois, 60208, USA}

    \date{\today}

    \begin{abstract}
        The series solution to Laplace's equation in a helical coordinate system is derived and
        refined using symmetry and chirality arguments.  These functions and their more commonplace
        counterparts are used to model solenoidal magnetic fields via linear, multidimensional
        curve-fitting.  A judicious choice of functional forms motivated by geometry, a small number
        of free parameters, and sparse input data can lead to highly accurate, fine-grained modeling
        of solenoidal magnetic fields. These models capture the helical features arising from the
        winding of the solenoid, with overall field accuracy at better than one part per million.
    \end{abstract}


    \maketitle

    \section{\label{sec:intro}Introduction}
\par
The use of superconducting solenoids is commonplace in scientific research, medical, and industrial
applications. The ever-increasing demand for higher resolution (e.g.\ medical imaging, charged
particle reconstruction) requires increased accuracy and precision for the magnetic
field~\cite{mu2e}. The physical geometry and helical winding of a solenoid generates an axial
asymmetry that may cause significant deviations of the magnetic field from that of an ideal
solenoid~\cite{NOURI201330, GU2011190}, and it may be necessary to account for these deviations.  The
deviations will be especially pertinent for solenoidal coils that have a relatively large pitch;
some large-pitch solenoids are currently employed in active particle physics experiments~\cite{L3,
alice}. A common analysis approach used within the high energy physics community is to combine a set
of sparse field measurements (e.g. provided by 2D or 3D Hall or NMR probes) with a series of
functions derived from solutions to Maxwell's equations to reconstruct the magnetic field within a
given volume~\cite{cms1, cms2, atlas}.  These methods have not attempted to use functions derived
with helical symmetry, however, which may be a better choice for modeling real solenoids.
\par
Any magnetic field must obey Maxwell's equations.  For a static field in a
region free of current and magnetic materials, the magnetic field $\bB$ can be expressed as $\bB=-\nabla\Phi$, where the scalar field $\Phi$ satisfies Laplace's equation:
\begin{equation} \label{eq:laplace}
    \nabla^2\Phi = 0.
\end{equation}
When treated as a boundary-value problem, \cref{eq:laplace} can
sometimes be solved via a separation of variables, and in the case of solenoids, this is typically
done in a cylindrical coordinate system~\cite{jackson_classical_1999}.  The axial symmetry inherent
to the cylindrical coordinate system is broken by the helical winding of the solenoid, however.
 In this case it is more appropriate to use a helical coordinate system, for which solutions to Laplace's equation via separation of variables are known to exist~\cite{overfelt}.
This coordinate system naturally encompasses the axial asymmetry of a helically wound solenoid, and
thus is a better choice for modeling the associated magnetic field.
\par
In this paper, we first derive the series solution to Laplace's equation in a helical coordinate
system.  We then calculate magnetic fields due to solenoids with different radius, pitch, and length
properties, and from those calculations produce datasets that are used as inputs for a parametric
modeling of the field. Using a judicious choice of terms for the helical series, cylindrical series, and other
physically valid functions for magnetic fields, we construct a model of the calculated magnetic fields using only a sparse
subset of the data and a linear sum of analytical functions whose coefficients are determined through least squares fitting.  We then compare the reconstructed field with a high-granularity dataset from the same calculated
field in order to determine the overall precision and accuracy of the model.
\par
This paper is structured as follows.  In \cref{sec:math}, we introduce the
mathematical formalism and solve Laplace's equation in a helical coordinate system.  In
\cref{sec:functions}, we establish the exact definitions of the functional forms used for modeling
magnetic fields in the following section. In \cref{sec:fit}, we describe the process used to
calculate helically-wound solenoids, and then model three specific examples using the functions
discussed in the previous sections.  For the third example, we investigate the impact of measurement error and data density on the model performance. We discuss the results and implications of these modeling
efforts in \cref{sec:conclusion}.

    \section{\label{sec:math}Mathematical Formalism}

% $$  \ddiff{\Phi}{x} + \ddiff{\Phi}{y} + \ddiff{\Phi}{z} = 0 $$ 

Representing a scalar potential function as an infinite linear combination of orthogonal basis
functions $\{\psi_n\}$ is a standard technique in potential theory, especially in service of solving
boundary value problems for Laplace's equation. 
\begin{equation*} \label{eq:series_phi}
    \Phi(x,y,z) = \sum_n c_n \psi_n(x,y,z).
\end{equation*}
\par
The convergence of this series is understood in the sense that $ \int |\Phi - \Phi_N|^2 dV
\rightarrow 0$, as $N\rightarrow\infty$, where $\Phi_N = \sum_{n=1}^N c_n \psi_n(x,y,z)$ is the
$N^{th}$ partial sum in the series representation of $\Phi$~\cite{evans10}. For special geometries, various
classical families of basis functions exist. For example, we will recall in what follows the
various forms of the harmonic functions which separate in cylindrical coordinates, and lead to
complete orthogonal bases of harmonic functions for certain function
spaces~\cite{jackson_classical_1999}. For our
application, it is not useful to have representations of potentials in an infinite expansion unless
we can well represent the target field using only a small number of terms. We are motivated to
choose the family of basis functions so that the above convergence happens as fast as possible.  
\par
Given field data $\bB(x_i,y_i,z_i)$ for a finite set of discrete points $(x_i, y_i, z_i)$, we propose a method to find a scalar potential function
$\Phi(x,y,z)$ which minimizes the sum of squared residuals:
\begin{equation}\label{eq:chi2}
    \chi^2 = \sum_i \nnorm{-\nabla\Phi(x_i,y_i,z_i) - \bB(x_i,y_i,z_i)}^2.
\end{equation}
We will use a series representation $\Phi =  \sum_j a_j\psi_j^{(cyl)} + \sum_j b_j\psi_j^{(hel)}$
in terms of a family of harmonic functions which we motivate below by the geometry of a solenoid
represented as a truncated helical curve. The cylindrical symmetry of a helix will cause us to work
in cylindrical coordinates $(r,\theta,z)$ at first, when we recall some facts about separable
cylindrical harmonics. Later we will work in a helical coordinate system introduced by
Waldron~\cite{waldron} and introduce a family of separable harmonic functions suited to the geometry
of any particular semi-idealized solenoid. Using this functional form for the scalar potential
$\Phi$ will allow us to model features of the magnetic field that are due to the helical nature of the
coil and the finite length of the solenoid.

%%%%%%%%%%%%%%%%%%%%%%
%%%%%%%%%%%%%%%%%%%%%%
\subsection{\label{subsec:cylhar}Cylindrical Harmonics}

We will briefly discuss the standard procedure of separation of variables of Laplace's equation in
cylindrical coordinates, in order to illuminate the connections to the following analogous work in
the helical coordinate system.
\par
In cylindrical coordinates $(r,\theta,z)$, given by the standard change of variables to Cartesian coordinates
$x=r\cos(\theta), y=r\sin(\theta),z=z$, Laplace's equation takes the following form:
\begin{equation*}
   \ddiff{\Phi}{r} +
   \frac{1}{r}\diff{\Phi}{r} +
   \frac{1}{r^2}\ddiff{\Phi}{\theta} +
   \ddiff{\Phi}{z}=0.
\end{equation*}
If we assume that $\Phi$ separates in the cylindrical variables, $\Phi(r,\theta,z) =
R(r)T(\theta)Z(z)$, we see that the functions $R(r), T(\theta), Z(z)$ must satisfy the following:
\begin{equation*}
    \frac{R''}{R} +
    \frac{1}{r}\frac{R'}{R} +
    \frac{1}{r^2}\frac{T''}{T} +
    \frac{Z''}{Z} = 0.
\end{equation*}
\par
The quantities ${Z''}/{Z}$ and ${T''}/{T}$ are constant
functions of the variables, so that $Z'' = c_z Z$, $T'' = c_\theta T$ for some real constants $c_z,
c_\theta$. The function $T(\theta)$ is $2\pi$-periodic,
which forces $c_\theta = -n^2$ for some integer $n$. Consequently, $R$ must satisfy the
following ordinary differential equation, which can be reduced to a Bessel equation of order $n$
using the linear change of variables $s = c r $, where $c=\sqrt{c_z}$:
\begin{equation*}
    \frac{R''}{R} + \frac{1}{r}\frac{R'}{R}  - \frac{n^2}{r^2} + c_z = 0.
\end{equation*}
\par
These relationships lead to the following forms for separable cylindrical harmonics depending on the sign of $c_z$. If $c_z = k^2$, then $Z(z) = A\cosh(kz) + B\sinh(kz)$ and $R(r) = CJ_n(kr)
+ DY_n(kr)$, where $J_n$ and $Y_n$ are the order $n$ Bessel functions of the first and second kind.
If $c_z = -k^2$, then $Z(z) = A\cos(kz) + B\sin(kz)$ and $R(r) = CI_n(kr) + DK_n(kr)$, where $I_n$
and $K_n$ are the order $n$ modified Bessel functions of the first and second kind.
\par
Because we are interested in potentials which do not have singularities at $r=0$, we discard the
Bessel functions of the second kind. Thus, for each choice of a positive integer $n$ and positive
real $k$, we have harmonic functions of the following form (note, for each term, there are
three other terms corresponding to replacing $\cos$ and $\cosh$ with $\sin$ and $\sinh$,
respectively):
\begin{equation*}
	\begin{split}
 	   \Phi_{n,k}(r,\theta,z) &= J_n(kr)\cos(n\theta)\cosh(kz) \\
  	  \Phi_{n,k}(r,\theta,z) &= I_n(kr)\cos(n\theta)\cos(kz). 
 	\end{split}
\end{equation*}
\par
This family of cylindrical harmonics is indexed by the integer $n$ and the real number $k$. When
interested in solving a boundary value problem on a cylinder $r<R$, $|z|<L$, one need only consider
the countable complete orthogonal basis corresponding to choosing $k$ to be $k_m = m\pi/L$, where $m$ is an integer.
\par
An alternative approach is to
assume that $T(\theta) = \exp(\omega_1 \theta)$ and $Z(z) = \exp(\omega_2 z)$ are complex
exponential functions parametrized by the complex separation constants $\omega_1, \omega_2$, and then determine the solution to the resulting ordinary differential equation for $R(r)$. In the case that the constants $\omega_1 = in$, and $\omega_2$ is real $\omega_2=k$ or purely imaginary
$\omega_2=ik$, this approach results in the same family of harmonic functions as the standard procedure. This alternative method is used by Waldron and Overfelt~\cite{waldron,
overfelt} to obtain separable solutions to the wave equation and Laplace's equation in the helical coordinate system.

% Fact: Idealized infinite right-handed solenoid with pitch 0 leads to uniform field inside where $\Phi(x,y,z) = Bz$ and $\bB = B\hat{z}$ 
\subsection{\label{subsec:heli_coords}Helical Coordinate Systems}

An idealized infinite helix is uniquely defined by the radius $R>0$, pitch\footnote{The
pitch, $P$, is defined as the advance along the axis of the helix for one complete turn.} $P>0$, and handedness ($+1$ or $-1$
corresponding to right-handed and left-handed helices, respectively). We use the convention that a
right-handed helix is one in which moving along the helix in the counter-clockwise direction (around
the $z$-axis when viewed from above) produces movement in the positive $z$ direction.
\par
When considering an idealized helix in space, it is convenient to work in cylindrical coordinates
and orient the helix so that it is contained in the cylinder given by the equation $r=R$, with the
$z$-axis serving as the axis of the helix. By rotating the helix (or the coordinate system),
we may assume that the helix intersects the $z=0$ plane on the positive $x$ axis, at the point
$(r,\theta,z) = (R,0,0)$.
\par
With these conventions, each right-handed helix may be parametrized by the following equations,
where $\pbar = P/2\pi$ and $t$ is a parameter representing the polar angle~$\theta$:
\begin{equation}
    \begin{split}\label{eq:ht}
	    \bh(t) &= ( x(t), y(t), z(t) ) = ( R\cos(t), R\sin(t), \pbar t )\\
          	    &= ( r(t), \theta(t), z(t) ) = ( R, t,  \pbar t  ).
    \end{split}
\end{equation}
\par
Any parametrization of a helix in Cartesian coordinates will require that all three
coordinates depend on the parameter $t$, and only by orienting the helix appropriately in a
cylindrical coordinate system can we parametrize the helix so that only two of the coordinates
depend on that parameter. For applications in which there is helical symmetry, it is convenient to
choose a coordinate system in which a helix can be parametrized using only one of the coordinates.

The following helical coordinate system $(r,\phi,\zeta)$ was introduced by Waldron~\cite{waldron}, and has many
similarities with a cylindrical coordinate system:
\begin{align*}
    x&= r \cos(\phi)		\quad & \quad	 r &= r\\
    y&= r \sin(\phi)		\quad & \quad	 \theta &= \phi	\\
    z&= \zeta + \pbar \phi	\quad & \quad	 z &= \zeta +\pbar \phi.
\end{align*}

\par
We need the Jacobian of the change of variables for the above coordinate transformations,
from which all of the identities concerning the coordinate directions, partial differential operators,
etc.\ are derived:

\begin{equation}\label{eq:jac}
    \dfrac{\partial(r,\theta,z)}{\partial(r,\phi,\zeta)} =
    \MAT{r_r & r_\ph & r_\z \\ \theta_r & \theta_\ph & \theta_\z \\ z_r & z_\ph & z_\z} =
    \begin{bmatrix} 1 & 0 & 0	\\ 	0 & 1 & 0	\\	0 & \pbar & 1  \end{bmatrix}.
\end{equation}

\par
\begin{figure*}[!htb]
    \includegraphics[width=\textwidth]{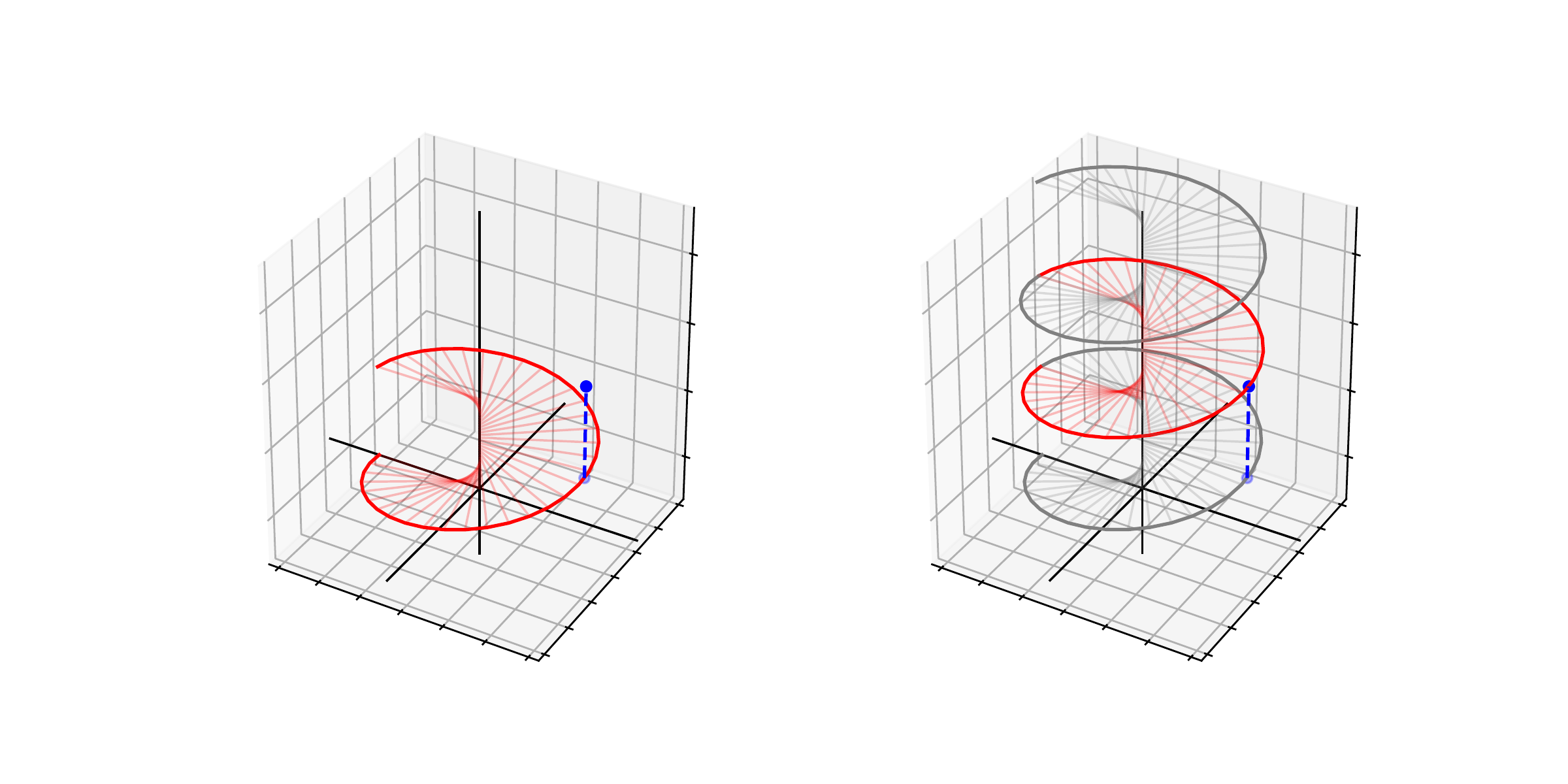}
    \centering
    \caption{Examples of a helical surface, swept out by the $r$ and $\phi$ variables, holding the
        $\zeta$ coordinate fixed. Equivalently, this is the surface defined by the helical equation
        $\zeta = 0$ in cylindrical which satisfies $z = \pbar\theta$. The blue point may be represented as
        $(r_0,\phi_0,\zeta_0+P)$ or as $(r_0,\phi_0+2\pi,\zeta_0)$, an example of the non-unique
        representation of points described in the text. (colour online) \label{fig:hel_surf}}
\end{figure*}

\Cref{fig:hel_coord} shows the three coordinate directions passing through a point
$(r_0,\phi_0,\zeta_0)$. The helical $r$ coordinate direction is identical with it's $r$
counterpart in cylindrical coordinates.  While the $\theta$ coordinate direction would wind around the cylinder
in a circle, the $\phi$ coordinate direction winds up the cylinder along a helical path. It is
important to note that although $\theta=\phi$ for each point in space, the vectors that point in the
$\theta$ and $\phi$ coordinate directions are different at each point. To be more explicit, the vector (in
cylindrical coordinates) which points in the $\phi$ coordinate direction is the second column in
\cref{eq:jac}, which has a component in the $\theta$ direction and a component in the $z$ direction.
Since the $\zeta$ coordinate differs from $z$ only by a translation, they are parallel: $\hat{z}=\hat{\zeta}$. The
$\zeta$ coordinate can be thought of as the distance in the $z$ direction between the point and the
surface $z=P\theta/2\pi$, illustrated by the length of the line segment \cref{fig:hel_surf}.

%%%%%%%%%%%%%%%%%%%%
\begin{center}
    \includegraphics[width=0.49\textwidth]{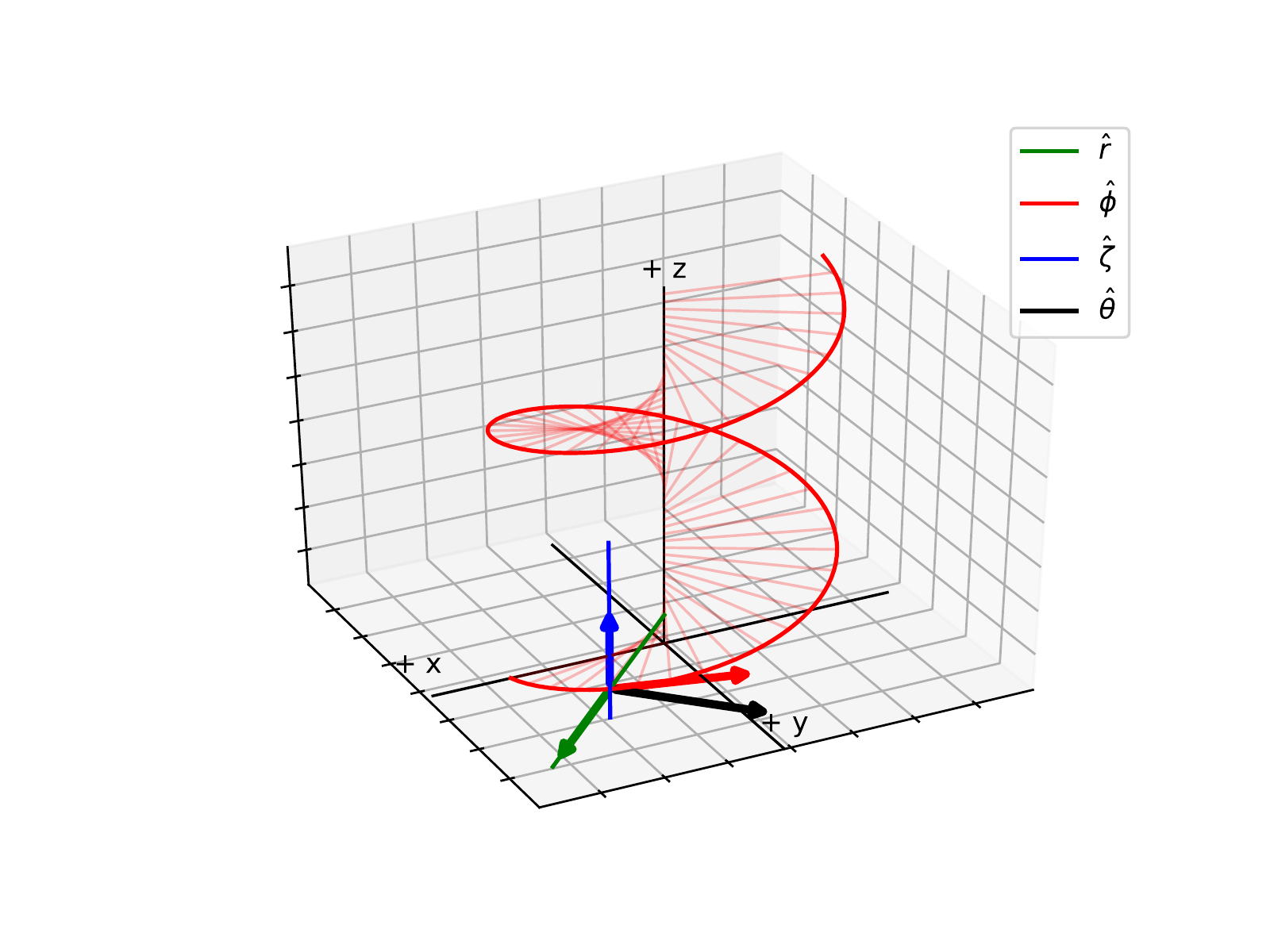}
    \captionof{figure}{Illustration of the three helical coordinate directions, $r$, $\phi$, and
        $\zeta$. The unit vectors in the coordinate directions $\hat{\phi}$,
        $\hat{\zeta}$, $\hat{\theta}$, and $\hat{z}$ all lie in the plane through the point which is tangent to
        the cylinder $r=r_0$ (i.e.\ perpendicular to $\hat{r}$). While $\hat{\theta}$ and $\hat{z}$ are
        orthogonal, $\hat{\phi}$ has a component in the $\hat{\theta}$ and $\hat{z}$ directions, and is
        thus not orthogonal to $\hat{z}$ or $\hat{\zeta}$. (colour online) \label{fig:hel_coord}}
\end{center}
%%%%%%%%%%%%%%%%%%%%

%\par
%The coordinate transformations declare that $r$ is identical to its counterpart in cylindrical. The
%$\phi$ variable can also be thought of as an angle, and although it always agrees numerically with
%the angle $\theta$ (up to a multiple of $2\pi$), there are some essential differences between
%$\theta$ and $\phi$ which are detailed below.

%%%%%%%%%%%%%%%%%%%%
%%%%%%%%%%%%%%%%%%%%

With these coordinates, we may parametrize a right-handed helix with radius $R$ and pitch $P$ as
follows:
\begin{equation*}
    ( r(t), \phi(t), \zeta(t) ) = ( R, t, 0 ).
\end{equation*}

\par
There is another difference between the cylindrical and helical coordinate system, relating
to non-unique representations of points. In cylindrical coordinates, each
point $(r,\theta,z)$ can be equivalently represented as $(r,\theta+2\pi k,z)$ for any integer $k$.
In helical coordinates the following more complicated identity holds for each point $(r,\phi,\zeta)$ and each integer $k$: 
\begin{equation*}
    (r,\phi+2\pi k,\zeta) = (r,\phi,\zeta + k\bP).
\end{equation*}
\par
This identity captures the fact that
$\Delta\phi = 2\pi k$ is equivalent to  $\Delta\zeta = k \bP$ (i.e., changes in the $\phi$ coordinate
which are integer multiples of $2\pi$ are equivalent to changes in the $\zeta$ coordinate which are
the same integer multiple of the pitch $\bP$). Informally, one gets to a point $k$ pitch-lengths
above by moving in only the $\phi$ coordinate or only in the $\zeta$ coordinate. This identity is
a consequence of the nonorthogonality of the $\phi$ and the $\zeta$ coordinate directions, represented by the
non-zero off-diagonal term in the Jacobian for the change of variables between cylindrical and
helical; relations like this are not possible in orthogonal coordinate systems.

% $$$$

%%%%%%%%%%%%%%%%%%%%%%
%%%%%%%%%%%%%%%%%%%%%%
\subsection{\label{subsec:lapace}Helical Harmonics}
\par
In Ref.~\cite{waldron}, Waldron introduces the right-handed helical coordinate system $(\hc)$ as a method for
analytically solving certain classic problems in electromagnetism in the case of helical symmetry.
He recasts Laplace's equation and the wave equation into helical coordinates, and then uses
a modified separation of variables technique to write down explicit analytic solutions to the
wave equation.
\par
Using the same technique, Overfelt~\cite{overfelt} solves Laplace's equation in all of space and
arrives at explicit forms for harmonic functions which separate in the helical coordinate system,
hereafter referred to as {\it helical harmonic functions}. Here we give an overview of the
technique, which parallels the method used for cylindrical coordinates.

\par
As $(r,\theta,z)$ are functions of $(r,\phi,\zeta)$ through the coordinate transformation equations, the
chain rule
\begin{equation*}
    \diff{f}{\phi} = \diff{f}{r}\diff{r}{\phi} + \diff{f}{\theta}\diff{\theta}{\phi} +
    \diff{f}{z}\diff{z}{\phi},
\end{equation*}
together with the appropriate entries of the Jacobian (\cref{eq:jac}), allow us to recover the
following relation among the differential operators in the coordinate directions:

\begin{align*}
	& \diff{}{r} = \diff{}{r}						\\
	& \diff{}{\phi} = \diff{}{\theta} + \pbar\diff{}{z}	\\
	& \diff{}{\zeta} = \diff{}{z}.
\end{align*}
Using these relations, we can translate Laplace's equation from
cylindrical coordinates into a helical coordinate system:
\begin{equation*}
    \ddiff{u}{r} + \frac{1}{r}\diff{u}{r} + \frac{1}{r^2}\bigg( \ddiff{u}{\phi} -2\pbar
    \mdiff{u}{\phi}{\zeta}+ \pbar^2 \ddiff{u}{\zeta} \bigg) = 0.
\end{equation*}
 If we look for separable solutions to Laplace's
equation of the form $R(r)P(\phi)Z(\zeta)$ where $P(\phi) = \exp(\omega_1\phi)$ and
$Z(\zeta)=\exp(\omega_2\zeta)$, then $R(r)$ satisfies the following differential equation:
\begin{equation}\label{eq:bessel1}
    r^2R''+rR'+r^2\omega_2^2R+(\omega_1-\pbar\omega_2)^2R=0.
\end{equation}

\Cref{eq:bessel1} also results from the choice $P(\phi) = \exp(-\omega_1\phi)$ and
$Z(\zeta)=\exp(-\omega_2\zeta)$. Similarly, by choosing $P(\phi) = \exp(\omega_1\phi)$ and
$Z(\zeta)=\exp(-\omega_2\zeta)$, or $P(\phi) = \exp(-\omega_1\phi)$ and
$Z(\zeta)=\exp(\omega_2\zeta)$, we find that $R(r)$ satifies
\begin{equation}\label{eq:bessel2}
    r^2R''+rR'+r^2\omega_2^2R+(\omega_1+\pbar\omega_2)^2R=0.
\end{equation}

As in the cylindrical case, the ordinary differential equations for $R$ can be transformed into Bessel equations of complex
order $i(\omega_1-\pbar\omega_2)$ in \cref{eq:bessel1} and $i(\omega_1+\pbar\omega_2)$ in \cref{eq:bessel2} in the variable $s$ through the linear change of variables $s =
\omega_2 r$. Thus, for arbitrary complex numbers, $\omega_1$ and $\omega_2$, we obtain the following
separable helical harmonics $\Phi_{\omega_1,\omega_2}(\hc) = R(r)P(\phi)Z(\zeta)$, in terms of Bessel functions of complex order and argument: 
\begin{align}
	\Phi_{\omega_1,\omega_2}(\hc) &= J_{i(\omega_1 - \pbar\omega_2)}(\omega_2 r)e^{\omega_1\phi}e^{\omega_2\zeta}		\label{eq:h1}\\
	\Phi_{\omega_1,\omega_2}(\hc) &= J_{i(\omega_1 - \pbar\omega_2)}(\omega_2 r)e^{-\omega_1\phi}e^{-\omega_2\zeta}	\label{eq:h2}\\
	\Phi_{\omega_1,\omega_2}(\hc) &= J_{i(\omega_1 + \pbar\omega_2)}(\omega_2 r)e^{\omega_1\phi}e^{-\omega_2\zeta}	\label{eq:h3}\\
	\Phi_{\omega_1,\omega_2}(\hc) &= J_{i(\omega_1 + \pbar\omega_2)}(\omega_2 r)e^{-\omega_1\phi}e^{\omega_2\zeta}.	\label{eq:h4}
\end{align}
\par
Although the helical harmonics separate in the variables $\hc$, when we add \cref{eq:h1,eq:h2}, the additive property of the
exponential function allows us to recover functions which couple the $\phi$ and $\zeta$ variables
inside the argument of trigonometric functions. Converting to cylindrical coordinates gives functions of the form
$R(r)F(\theta, z)$ which does not separate in the cylindrical variables $r,\theta,z$.

\par
In particular, choosing $\omega_1 = in^*$ and $\omega_2 = ik$ in \cref{eq:h1,eq:h2}, where $n^*$ is an integer and $k$ is a real constant, expressing the complex exponentials in terms of trigonometric functions and the Bessel functions in terms of modified Bessel functions, we arrive at the following form of
real-valued helical harmonics, which we also represent in cylindrical coordinates and refer to as ``left-handed" helical harmonics:
\begin{align*}
    \Phi_{n,k}(r,\phi,\zeta) &= I_{n^*-\pbar k}(kr)\cos(n^*\phi + k\zeta)		\\
    \Phi_{n,k}(r,\theta,z) &= I_{n^*-\pbar k}(kr)\cos((n^*-\pbar k)\theta + kz).
\end{align*}
The same choice of $\omega_1 = in^*$ and $\omega_2 = ik$ in \cref{eq:h3,eq:h4} results in the following ``right-handed" helical harmonics:
\begin{align*}
    \Phi_{n,k}(r,\phi,\zeta) &= I_{n^*+\pbar k}(kr)\cos(n^*\phi - k\zeta)		\\
    \Phi_{n,k}(r,\theta,z) &= I_{n^*+\pbar k}(kr)\cos((n^*+\pbar k)\theta - kz).
\end{align*}		
\par
Again, we have a family of harmonic functions indexed by a real parameter $k$. For notational ease,
we are free to set $n=n^*-\pbar k$, and since $k$ controls the oscillations of
$\Phi$ with respect to $z$, by setting $k$ to be integer multiples of the effective pitch, $k_m = m/\pbar$, we recover a countable family $\{\Phi_{n,k_m}\}$ of helical harmonic
functions which respect the periodicity of the solenoidal coil in $z$.
\par
On any cylinder $r=r_0$, each right-handed helical harmonic function takes the following form,
where $R(r)$ is a modified Bessel function:
\begin{equation}\label{eq:hel_harmonic}
    R(r_0) \left(A_{n,m}\cos(n\theta - k_mz)+B_{n,m}\sin(n\theta - k_mz)\right).
\end{equation}
\par
\begin{figure*}[!htb]
    \includegraphics[width=0.32\textwidth]{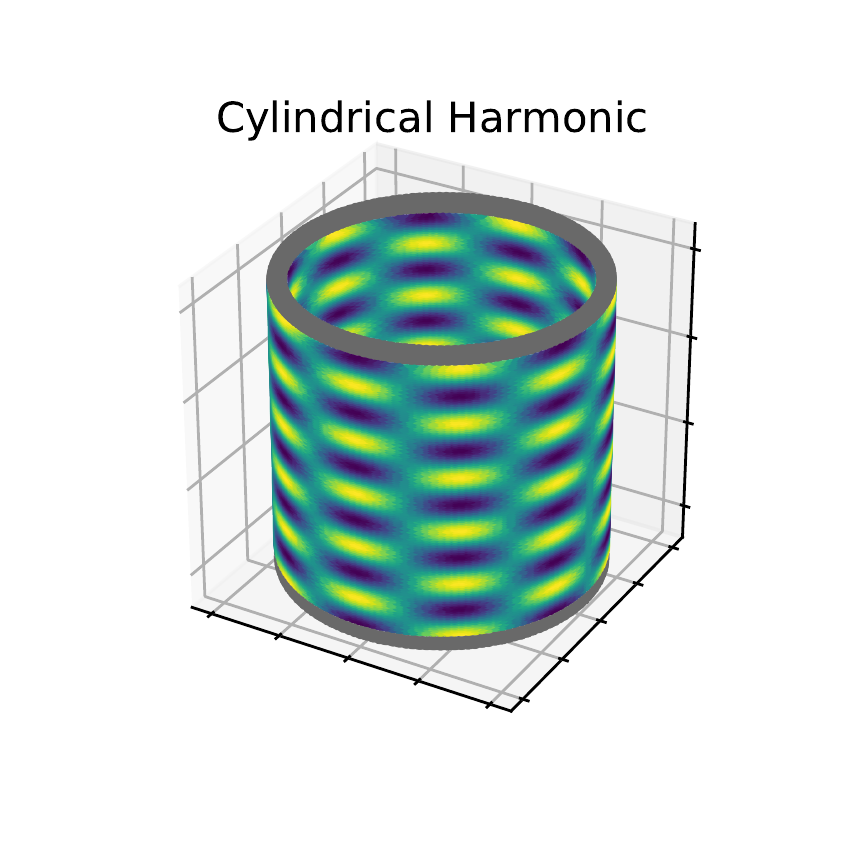}
    \includegraphics[width=0.32\textwidth]{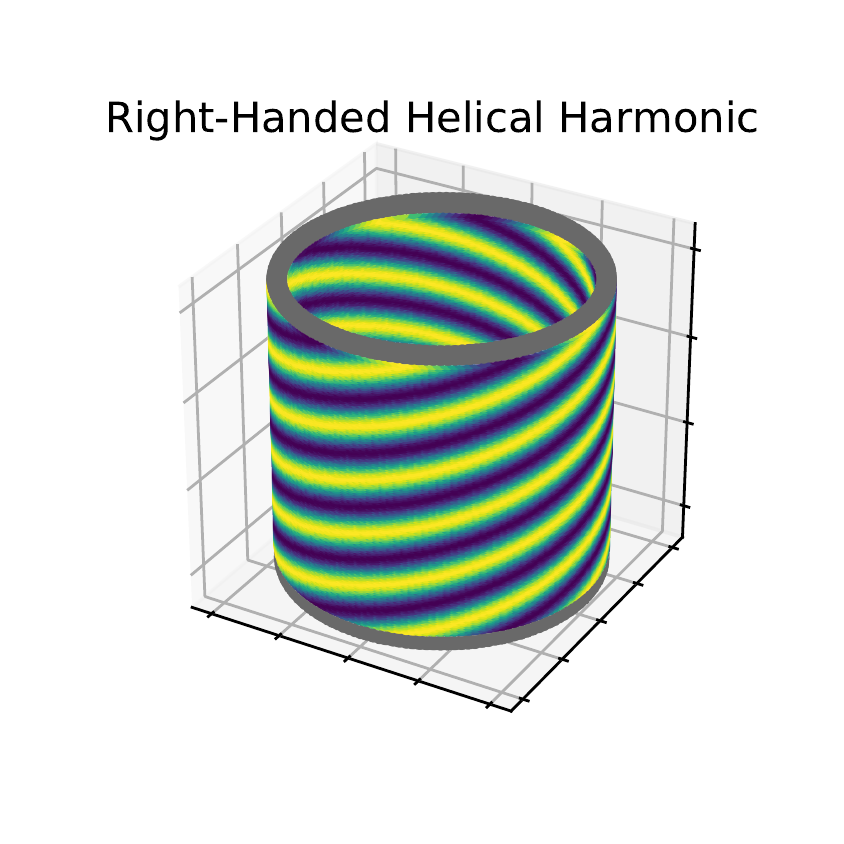}
    \includegraphics[width=0.32\textwidth]{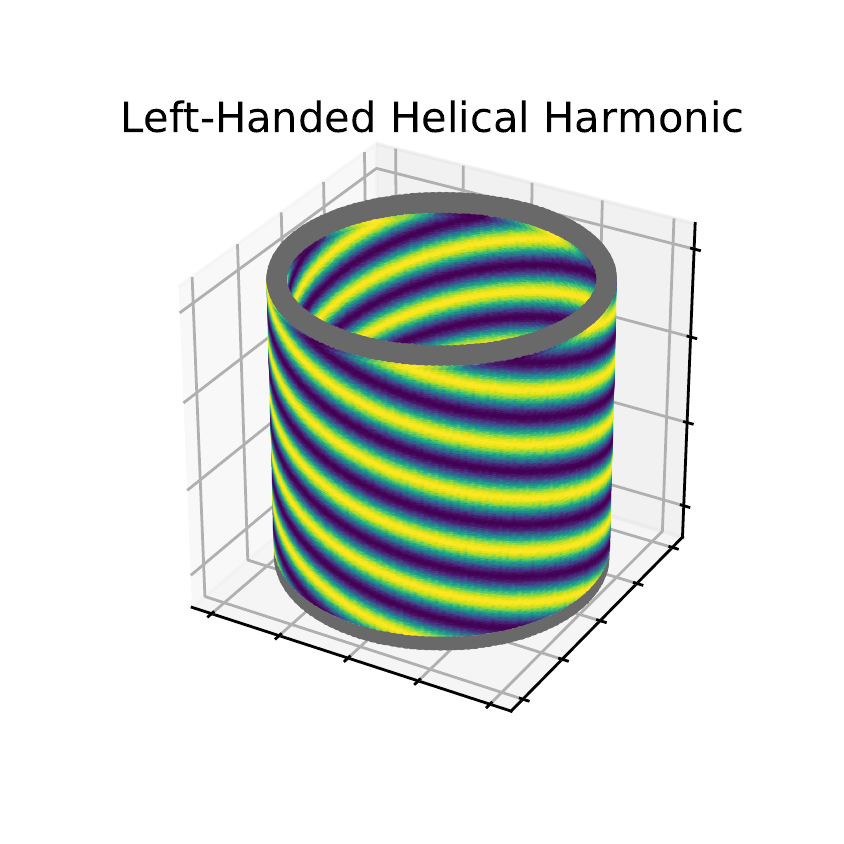}
    \centering
    \caption{Examples of a scalar field due to a single harmonic function $\Phi_{5,5}(r,\theta,z)$
    restricted to a fixed cylinder $r=$ const. The dark points indicate where $\Phi$ is positive,
while the light points indicate where $\Phi$ is negative. Note that the right-handed (left-handed)
helical harmonic function is constant along right-handed (left-handed) helices, while the
cylindrical harmonic function is not constant along any helix. (colour online)}\label{fig:scalar_surfs}
\end{figure*}

 
The expression in \cref{eq:hel_harmonic} is constant along the curves in the cylinder $r=r_0$ that satisfy
$n\theta - k_mz =$ const. That is, it is constant along curves $z = n\theta/k_m +$
const, which are all the right-handed helices with pitch $P^* = 2\pi n/k_m$. If we work in a
helical coordinate system with pitch $P$, we can make $P^*=P$ by choosing $k_m = 2\pi m/P$ and
letting $n=m$. It is also true that a left-handed helical harmonic function is constant along certain left-handed helices, whose pitch is determined
as above. This tells us that the family $\{\Phi_{n,k_m}\}$ introduced above contains a sequence of
terms $\Phi_{n,k_n}$ which are all constant along right-handed helices with pitch $P$. In
\cref{fig:scalar_surfs} we show cylindrical and helical harmonic functions restricted to a cylinder $r=r_0$.

If the field $\bB$ is due to an idealized wire, parametrized by $\bh(t)$ (\cref{eq:ht}),
carrying a uniform current $I$, then the field is given by the Biot-Savart Law:
\begin{equation}\label{eq:bs}
    \bB(\bx)=\frac{\mu_0}{4\pi} \int \frac{Id\bh\times(\bx-\bh)}{|\bx-\bh|^3}.
\end{equation}
\par

{\bf Theorem 1:} A field $\bB$ given by the Biot-Savart Law for an idealized infinite right-handed
helix with pitch $P$ leads to a field which is cylindrically symmetric along right-handed helical
curves with pitch $P$, that is $\bB(r_0,\theta_0,z_0) = \bB(r_0,\theta_0+2\pi s,z_0+Ps)$ for all $(r_0,\theta_0,z_0)$ and all $s$, when $\bB$
is represented in cylindrical coordinates $\bB = B_r\hat{r} + B_\theta\hat{\theta}+ B_z\hat{z}$.
\par
\begin{proof}
For each real $s$, the change of coordinates $C_s(r,\theta,z) = (r,\theta+2\pi s, z + Ps)$ is an
orientation preserving rigid motion of space under which infinite right-handed helices of pitch $P$
are invariant. $C_s$ is a translation of the cylindrical coordinate system along a helix with
pitch $P$, and is the composition of a rotation about the $z$-axis by an angle $2\pi s$ and a
translation in the $z$ direction with $\Delta z=Ps$. This isometry $C_s$ also acts on vectors via
the Jacobian $JC_s = \frac{\partial C_s}{\partial(r,\theta,z)}$, sending a vector $\bv_0$ at $\bx_0$
represented in cylindrical components, to the vector $\bv_1 = JC_s(\bx_0)\bv_0$ at $\bx_1=C_s(\bx_0)$. Since $C_s$ is a
translation in cylindrical coordinates, the Jacobian is the identity matrix, and so the vectors
$\bv_0$ and $\bv_1$ have the same cylindrical components.  Now, since this rigid motion leaves the
geometry of the current carrying device unchanged, since $C_s(\bh(t)) = \bh(t+s)$, the resulting
magnetic field must be invariant under the transformation $C_s$: $\bB(C_s(\bx_0)) =
JC_s(\bx_0)\bB(\bx_0) = \bB(\bx_0)$, and the result follows.
\end{proof}
\par
%Theorem 1 motivates using harmonic functions which are constant along helical curves with pitch $P$

\par
In the special case when $s$ is an integer, the translation in $\theta$ is an integer
multiple of $2\pi$, so the transformation $C_s$ is equivalent to a pure translation in the $z$
direction with $\Delta z = sP$, an integer multiple of the pitch. In this case, the result of
Theorem 1 is equivalent to the statement that $\bB$ is $P$-periodic in $z$. Theorem 1
implies the more general statement that in any parallel planes $z = c_1$ and $z = c_2$, the field
$\bB$ is identical, up to a rotation by $\Delta \theta = 2\pi(c_2-c_1)/P$.

\par
Theorem 1, together with the preceding properties of the real-valued helical harmonics, leads to our
choice of right-handed helical harmonic functions to model a field due to a right-handed helical
solenoid, since these are precisely the harmonic functions which are constant along curves with the
handedness and pitch of the solenoid.

    \section{\label{sec:functions}Functional Forms}
\par
This section details the individual functional forms used during the fitting process.  Functions are
expressed in either cylindrical or Cartesian coordinates, as these coordinate systems are typically used when measuring magnetic field points in solenoids, and allow for more straightforward comparisons
between different forms.  Functional forms are broken into three classes: ``cylindrical'',
``helical'', and ``additional.'' These classes differentiate between functions derived from 
series solutions to Laplace's equation in cylindrical coordinates, series solutions to Laplace's
equation in helical coordinates, and additional functions that help with specific sources of
magnetic field asymmetry. 

\subsection{\label{sec:cyl_func}Cylindrical Functions}
Magnetic fields stemming from idealized solenoids exhibit monotonic behavior for all field
components as a function of $r$.  This motivates the use of modified Bessel functions as a general
solution to solenoidal magnetic fields, as modified Bessel functions exhibit monotonic behavior,
while regular Bessel functions exhibit oscillatory behavior.  From that equation, we can implement a
fully real-valued series expression using $2(m\cdot n)+n$ free parameters:
\begin{equation}\label{eq:scalar_cyl}
    \begin{split}
        \Phi(r,\theta,z) =& \sum_{m,n}I_n(k_{m}r)[C_n\sin(n\theta)+(1-C_n)\cos(n\theta)]\cdot\\
        &[A_{m,n}\cos(k_{m}z)+B_{m,n}\sin(k_{m}z)].
    \end{split}
\end{equation}
The coefficients $A_{m,n}$, $B_{m,n}$, and $C_{n}$ are determined by the fitting procedure described
in \cref{sec:fit}.  The quantity $k_{m}$ is equal to $(m+1)\pi/L$, where $L$ is some effective length scale and $m+1$ is used instead of $m$ in order to avoid the trivial constant
term which is better handled as described in \cref{sec:add_func}.
\par
Physically, the scalar potential $\Phi(r,\theta,z)$ cannot be observed or measured directly.
Instead, a least-squares fit to the magnetic field components is used to determine the free parameters $A_{m,n}, B_{m,n}, C_n$:

%{\red perhaps write down $\chi^2=\sum_j (\bB_i^{meas}(x_j) - \bB_i^{expan}(x_j) )^2$ }

\begin{align}\label{eq:triple_cyl}
    \begin{split}
        B_r =& \sum_{m,n}I'_n(k_{m}r)[C_n\sin(n\theta)+(1-C_n)\cos(n\theta)]\cdot\\
        &[A_{m,n}\cos(k_{m}z)+B_{m,n}\sin(k_{m}z)]k_m\\
        B_{\theta} =& \sum_{m,n}I_n(k_{m}r)[C_n\cos(n\theta)-(1-C_n)\sin(n\theta)]\cdot\\
        &[A_{m,n}\cos(k_{m}z)+B_{m,n}\sin(k_{m}z)]\frac{n}{r}\\
        B_z =& \sum_{m,n}I_n(k_{m}r)[C_n\sin(n\theta)+(1-C_n)\cos(n\theta)]\cdot\\
        &[-A_{m,n}\sin(k_{m}z)+B_{m,n}\cos(k_{m}z)]k_{m},
    \end{split}
\end{align}
where $I'_m = \frac{d}{dr}\left[ I_m \right]$ is the ordinary derivative of the modified Bessel function.
%The scalar potential, $\Phi(r,\theta,z)$, cannot be used explicitly in a real-world example.
%Only magnetic fields, not magnetic potentials, can be directly measured.  Therefore, the gradient of
%the scalar potential, $-\nabla\Phi = \bB$, is used to directly model a magnetic field.  
\par
The expressions in \cref{eq:triple_cyl} can be carried out to any number of terms in $m$ and $n$.
In practice, the truncation of these series is determined by the required modeling accuracy.  In general, more asymmetric features in the $z$-direction necessitate a larger number of
$m$ terms, while more axial asymmetry necessitates a larger number of $n$ terms.

\subsection{\label{sec:hel_func}Helical Functions}
As expected from the cylindrical functions, there is no direct
coupling between the $\theta$ and $z$ variables in \cref{eq:scalar_cyl,eq:triple_cyl}.  In order to account for axial asymmetries
induced by the helical coil, one would have to expand the cylindrical series to a very large
number of terms.  In contrast, the helical harmonic functions link the $\theta$ and $z$ coordinates, and choosing
the appropriate pitch parameter will lead to a large reduction in free parameters needed to obtain an accurate modeling of the magnetic field.  As described in \cref{subsec:lapace}, a
right-handed helix leads to surfaces of constant potential via a right-handed series solution.
The following examples are based solely on right-handed coils, so only the right-handed
terms are used for the helical series expansion. From \cref{eq:hel_harmonic}, the scalar field can be expressed in the following manner:
\begin{equation}\label{eq:scalar_hel}
    \begin{split}
        \Phi(r,\theta,z) =
        \sum_{m,n}I_{n}\left(h_m r\right)
        [ & D_{m,n}\cos(n\theta-h_m z)+ \\
          & E_{m,n}\sin(n\theta-h_m z)],
    \end{split}
\end{equation}
where $h_m = (m+1)/\pbar$.  
As before, we use $m+1$ instead of $m$ to avoid the trivial constant term.  
The terms $D_{m,n}$ and $E_{m,n}$ are constants.  We obtain the expressions for the magnetic field components:
\begin{align}\label{eq:triple_hel}
    \begin{split}
        B_r =& \sum_{m,n}I'_{n}\left(h_m r\right)
        [ D_{m,n}\cos(n\theta-h_m z)+ \\
        & E_{m,n}\sin(n\theta-h_m z)]h_{m}\\
        B_{\theta} =& \sum_{m,n}I_{n}\left(h_m r\right)
        [ -D_{m,n}\sin(n\theta-h_m z)+ \\
        & E_{m,n}\cos(n\theta-h_m z)]\frac{n}{r}\\
        B_{z} =& \sum_{m,n}I_{n}\left(h_m r\right)
        [ D_{m,n}\sin(n\theta-h_m z)- \\
        & E_{m,n}\cos(n\theta-h_m z)]h_m.
    \end{split}
\end{align}

\subsection{\label{sec:add_func}Additional Functions}
\par
The helical and cylindrical harmonic series expansions are constructed in the context of an infinitely long solenoid.  Therefore, they are not equipped to handle constant fields, nor the edge effects from a finite solenoid; two types of behavior we expect to observe.
\par
Constant fields can be handled simply by using a first-order Cartesian solution to Laplace’s equation. For the examples used in this paper, the solenoid axis runs along the $z$-axis, and there is a near-constant $B_z$ field component.  This implies that a $\Phi(x,y,z) = cz$ term in the potential expansion can represent that field appropriately. In practice, by including the $cz$ term we are able to reduce the total number of free parameters needed to accurately model the magnetic field.  In particular, this term also replaces the zeroth order Bessel function terms ($I_0(r)$), which are the only terms capable of producing non-zero values for the magnetic field at $r=0$. 
\par
The final set of terms used in the following fitting procedure are derived directly from the
Biot-Savart Law (\cref{eq:bs}).  An infinitesimal current located at $(x_0,y_0,z_0)$ element can be modeled using the following form:
\begin{equation*}\label{eq:bs_terms}
    \bB(x,y,z) = \frac{{\bf F}\times{\bf r}}{{\bf r}^3},
\end{equation*}
where ${\bf F} = (F_x, F_y, F_z)$ are the free parameters associated with an effective infinitesimal
current element, and ${\bf r} = (x-x_0, y-y_0, z-z_0)$.  These terms are included to account for small asymmetric contributions arising from the initial and terminal endpoints of the solenoid simulations, detailed in the following section.  Two of these infinitesimal current element terms are included during fitting, with the spatial free parameters ($x_0, y_0, z_0$) constrained to resided within $1\,\textrm{cm}$ of the respective endpoints.

%{\red The edge-effects due to the finite size of the solenoid must be handled via a different set of terms.  These effects manifest themselves as an axial asymmetry without any direct dependence on the pitch or winding of the solenoid, see Fig.~\ref{fig:edge_effects}.  This asymmetry is minimized for points sufficiently far from the edges of the solenoid, but cannot be ignored in most instances. The exact locations of the beginning and ending of the solenoid have a direct impact on the direction of the imposed asymmetry (Fig~\ref{fig:quarter_turn}).  Given these features, it was found that a term inspired by an infinitesimal current element was sufficient for modeling this aspect of the magnetic field.  This Biot-Savart term is implemented as follows:}

    \section{\label{sec:fit}Calculation and Modeling}
\subsection{\label{sec:matlab}Calculating  a Helical Solenoid}
\par
\begin{figure}
    \includegraphics[width=0.49\textwidth]{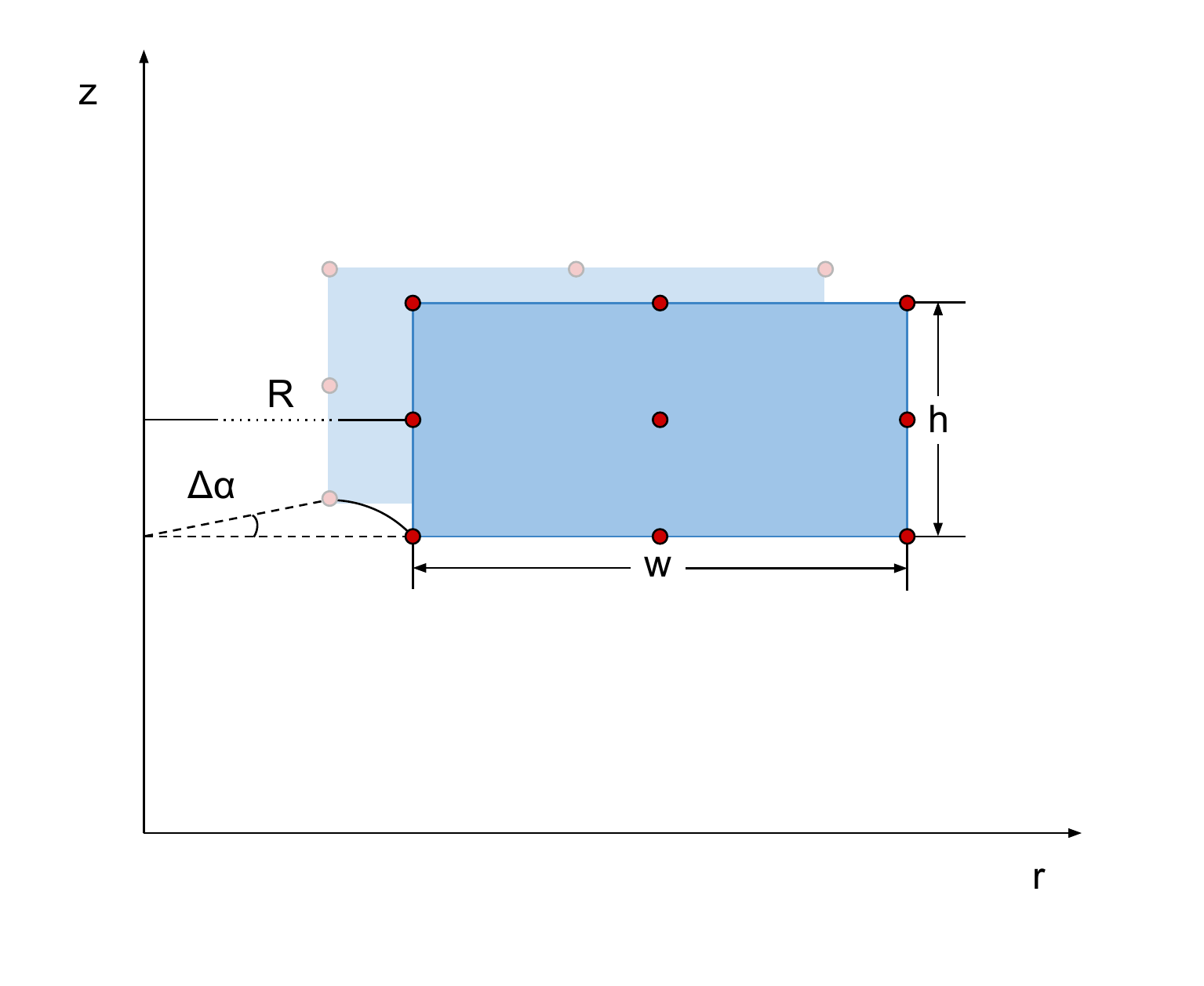}
    \centering
    \captionof{figure}{The solenoid ribbon cable cross section used for simulation, in the
        $rz$-plane.  $h$ and $w$ are the height and width of the cable, $R$ is the radius of the
        solenoid, and $\Delta\alpha$ is the angle between adjacent planes.  $\Delta\alpha$ winds
        along a direction of constant helix at a given pitch. (colour online)}\label{fig:cable_xs}
\end{figure}

In order to determine the benefit of a helical solution to Laplace's equation, we simulate magnetic
fields due to helically-wound solenoids.  Using MATLAB~\cite{MATLAB}, we define three helical coil
geometries which would be typical of high-field, large-bore solenoids used in particle physics
detectors.  The coils consist of flat, ribbon-like cables (\cref{fig:cable_xs}) with a constant and
homogeneous current and without any additional leads or connections.  The cables have a width of
5.25\,mm and a height of 2.34\,mm.  The magnetic fields they generate are calculated through use of the
Biot-Savart law, in which a 3D, parallelized numerical integral is computed to obtain the magnetic
field at a grid of points inside the coil.  When performing the integral, the angular step size, $\Delta\alpha$, between cross-sectional planes of the cable is roughly $5.8\cdot10^{-4}$\,radians.   The parameters
used for the different fields generated for this study are shown in \cref{tab:coil_params}.

%\begin{table*}
%    \begin{center}
%        \begin{tabular}{>{\centering\arraybackslash}m{2cm} >{\centering\arraybackslash}m{2cm}
%            >{\centering\arraybackslash}m{2cm} >{\centering\arraybackslash}m{2cm}}
%            \toprule
%            Parameter & Solenoid A& Solenoid B& Solenoid C\arraybackslash\\
%            \midrule
%            \rowcolor{black!20} Length & 92m & 9.2m & 9.2m \\
%            Radius & 250mm & 1000mm & 250mm \\
%            \rowcolor{black!20} Pitch & 10cm & 7.5mm & 10cm \\
%            Current & 6114 A  & 6114 A & 6114 A \\
%            \rowcolor{black!20} Average $B_z$ & 767 G & 1 T & 767 G \\
%            \bottomrule
%        \end{tabular}
%    \end{center}
%    \caption{Table of simulation parameters for different solenoid examples.}\label{tab:coil_params}
%\end{table*}

\begin{table*}[!htbp]
    \begin{center}
            \begin{tabular}{| c | c | c | c |}
            	\hline
            	~Parameter~ & ~Solenoid A~& ~Solenoid B~& ~Solenoid C~ \\\hline
		Length & 92\,m & 9.2\,m & 9.2\,m \\\hline
            	Radius & 250\,mm & 1000\,mm & 250\,mm \\\hline
		Pitch & 10\,cm & 7.5\,mm & 10\,cm \\\hline
            	Current & 6114\,A  & 6114\,A & 6114\,A \\\hline
		Average $B_z$ & 767\,G & 1\,T & 767\,G \\\hline
        \end{tabular}
    \end{center}
    \caption{Table of simulation parameters for different solenoid examples.}\label{tab:coil_params}
\end{table*}
\par
The geometries of the three different solenoids are chosen to probe different magnetic field
features.  Solenoid A is long (92\,m) compared to its radius (25\,cm).  It also has
a large pitch (10\,cm), which emphasizes the helical field features.  These properties lead to a
field that exhibits ideal helical behavior in the center region of the bore, with little impact
from edge effects.  Solenoid B is both shorter (9.2\,m) and has a larger radius (1\,m) than solenoid~A.
It also has a smaller pitch (7.5\,mm), which reduces the helical field features to a negligible degree.  This field can be modeled with cylindrical terms only.  Solenoid C has geometric properties
from both A and B.  It has the same radius and pitch as Solenoid A, but with a length equal to
Solenoid B.  To model this field appropriately, we use a combination of helical and cylindrical
terms.
\par
\begin{figure*}[!htb]
    \includegraphics[width=0.49\textwidth]{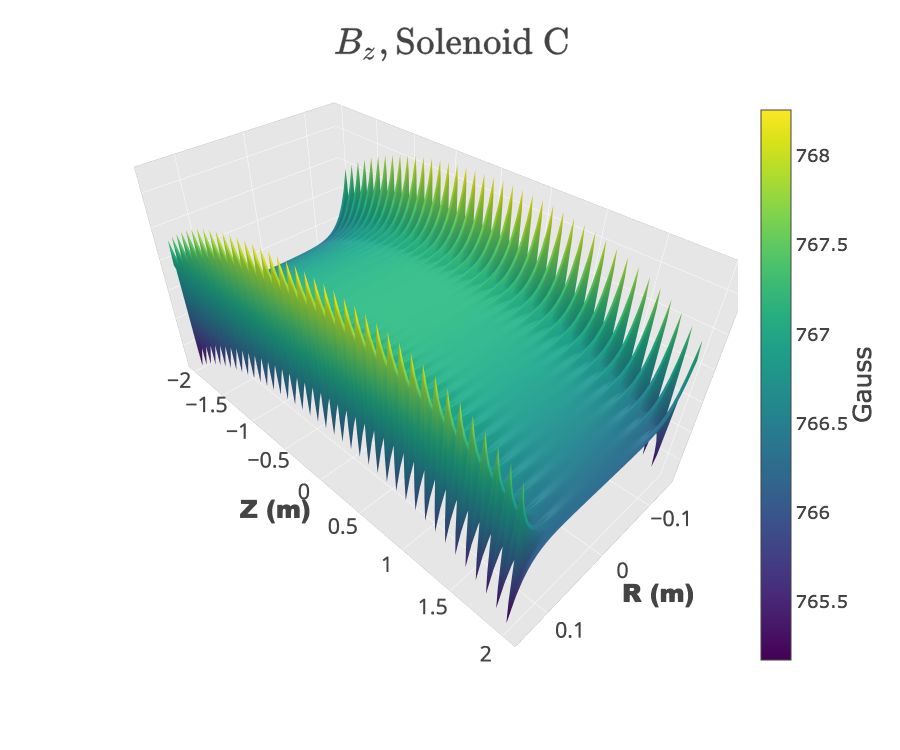}
    \includegraphics[width=0.49\textwidth]{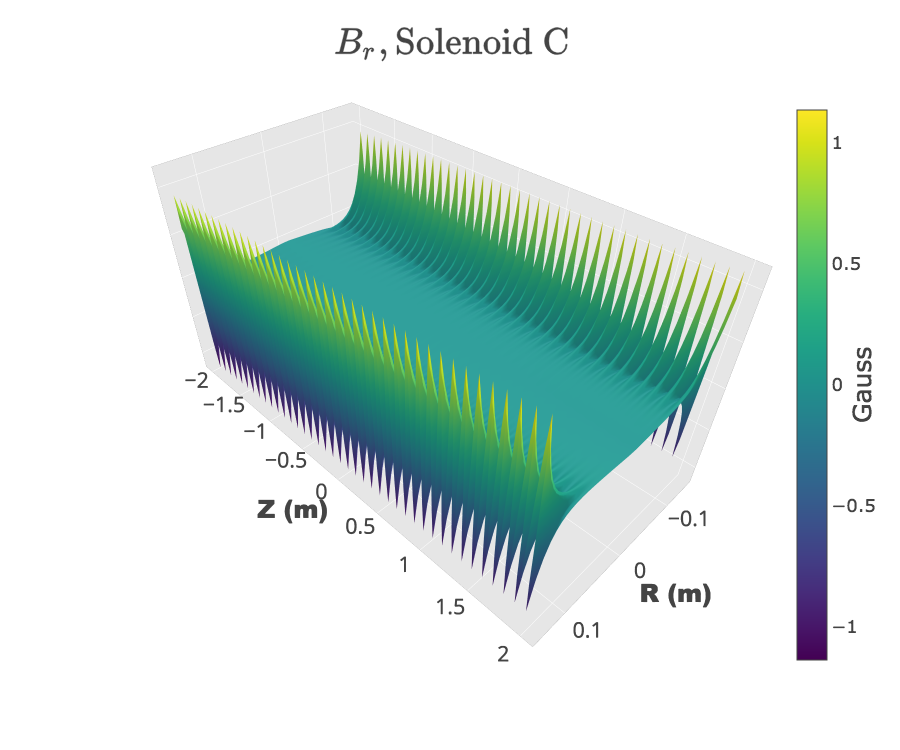}
    \includegraphics[width=0.49\textwidth]{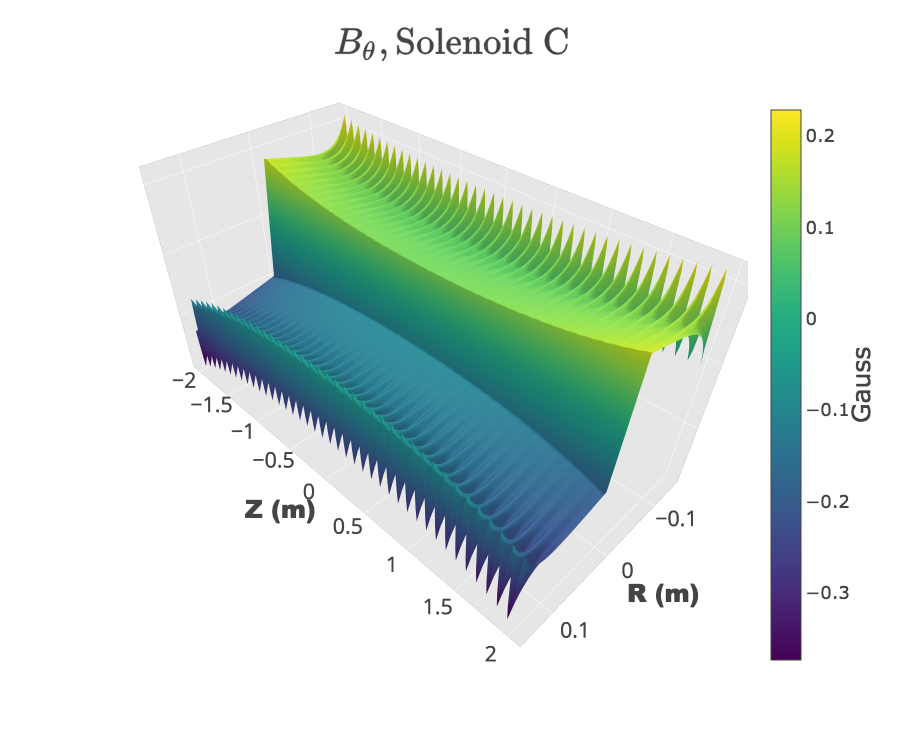}
    \centering
    \caption{Magnetic field components, $B_z, B_r, B_{\theta}$, using 2D slices of
        Solenoid C at a fixed angle $\theta$. (colour online)}\label{fig:mag_field_ex}
\end{figure*}

Depictions of the field components for Solenoid C are given in \cref{fig:mag_field_ex}.
These magnetic fields have a relatively large, constant field in the $z$-direction, as is characteristic of
solenoids.  The radial and axial components are orders of magnitude smaller than the $z$-component (as
long as one is far from the edges of the solenoid), but they are not negligible.
Due to the helical nature of the solenoid, all field components exhibit
wiggles in the $z$-direction, with a wavelength given by the pitch of the solenoid.  

\subsection{\label{sec:fitting}Fitting Magnetic Field Data}
The magnetic fields can be generated
using as fine a grid as required (or that computing power allows).  However, a real-world solenoid
cannot be surveyed on an arbitrarily fine grid.  Instead, a solenoid is typically surveyed
sparsely, and from that survey, the full magnetic field is reconstructed.  In this
section, we use the harmonic functions derived above to fit a sparse set
of magnetic field values, and reconstruct a continuous field using those functions.
Judicious choices of the number of parameters and length scales allow us to model
a range of magnetic fields with a high degree of accuracy.
\par The free parameters are determined by fitting $(B_r, B_\theta, B_z)$ from a subset of the
grid points.  The remaining grid points are
used for validation and visualization.  The sparse subset of grid points used for fitting is selected by mimicking the process of surveying a real-world solenoid using a mechanical field mapper~\cite{mu2e}.  This results in a cylindrically symmetric grid with constant spacing in $r, \theta, z$, the exact values of which are detailed in the following subsections. %{\blue \st{The sparse subset of grid points used for fitting is selected in the following manner: A set of cylindrical radial distances are chosen, as measured from the center line of the solenoid.  These distances are contained within the bore of the solenoid.   A set of azimuthal angles are chosen; the radial positions will sweep out circles in the axial direction using a fixed step-size.  A set of positions along the axis of the solenoid is chosen so that a 3D set of cylindrically symmetric grid points is obtained.}}
\par
These grid points and their associated magnetic field components, $\bB(x_i, y_i, z_i)$, are used as
independent variables for the fit.  The sum of squared residuals, $\chi^2$~(\cref{eq:chi2}),
is the objective function which is minimized using least-squares optimization algorithms from the
Scipy and lmfit packages for the Python software language~\cite{scipy,lm_algo, trf_algo, lmfit}, and
linear least-squares solver in MATLAB~\cite{MATLAB}.
\par
In addition to either the partial derivatives of the cylindrical (\cref{eq:scalar_cyl}) and/or helical
(\cref{eq:scalar_hel}) series expansions, the additional terms discussed in \cref{sec:add_func}
are included with all fitting functions.  
\par
To determine the optimal number of terms in a given series expansion, fits are performed
iteratively, increasing the number of $m$ and $n$ terms at each iteration.  Once
the desired accuracy is reached, no additional terms are added, even if additional terms leads to
slight improvements with respect to the quality of fit.  The accuracy requirements for the three
solenoid examples are discussed in more detail in the following sections.
\par
Additional studies are performed on Solenoid C, in order to investigate the impact of measurement error and grid density on the fit performance.  Gaussian noise is introduced to the sparse field component values to emulate measurement error.  The density of the sparse grid
is varied, reducing the total number of data available during fitting.  

\subsubsection{\label{sec:solA}Solenoid A}
\par
Solenoid A (\cref{tab:coil_params}) is intended to bring out the helical features that arise
from the winding of the solenoid, while minimizing edge effects.  The solenoid is very long
with respect to its radius, has a relatively large pitch, and region of interest is restricted to the
middle of the solenoid, far from the edges.  This region is fit with two different series
expansions: the cylindrical expansion (\cref{eq:triple_cyl}) and the helical expansion
(\cref{eq:triple_hel}) in order to ascertain which expansion leads to greater accuracy and fewer
terms.  The two expansions are compared using the objective function (\cref{eq:chi2}), the
per-component residuals, and the overall accuracy using the validation data set.
\par
The sparse set of data selected to fit for Solenoid A consists of 6 radial positions, chosen such
that $12.5\,\textrm{mm} < r < 150\,\textrm{mm}$, 16 equally-spaced angles from 0 to 2$\pi$, and 60
equally-spaced $z$ positions $-1500\,\textrm{mm} < z < 1500\,\textrm{mm}$. The total number of data
points is 5760, and each point contains three magnetic field vector components, $B_r, B_{\theta},
B_z$.  The validation dataset is generated from a selection of 24 equally spaced radial positions
from $6.25\,\textrm{mm} < r < 150\,\textrm{mm}$, 126 equally spaced angles, and 120 equally spaced Z
positions, leading to a total of 362,880 data points.  These points were chosen to probe a region far
from the edges of the solenoid to minimize edge effects, and far enough from the solenoid coil to
avoid discontinuities while still capturing the helical wiggles.  The pitch, $P$, chosen for the
helical series is 10\,cm, identical to the pitch used to generate the solenoid.  The length scale,
$L$, used for the cylindrical series is 5\,cm.  This length scale was chosen such that $k_m =h_m$ for
\cref{eq:triple_cyl,eq:triple_hel}, which determines the wavelength of the wiggles in the
$z$-direction.
\par
\begin{figure*}
    \includegraphics[width=\textwidth]{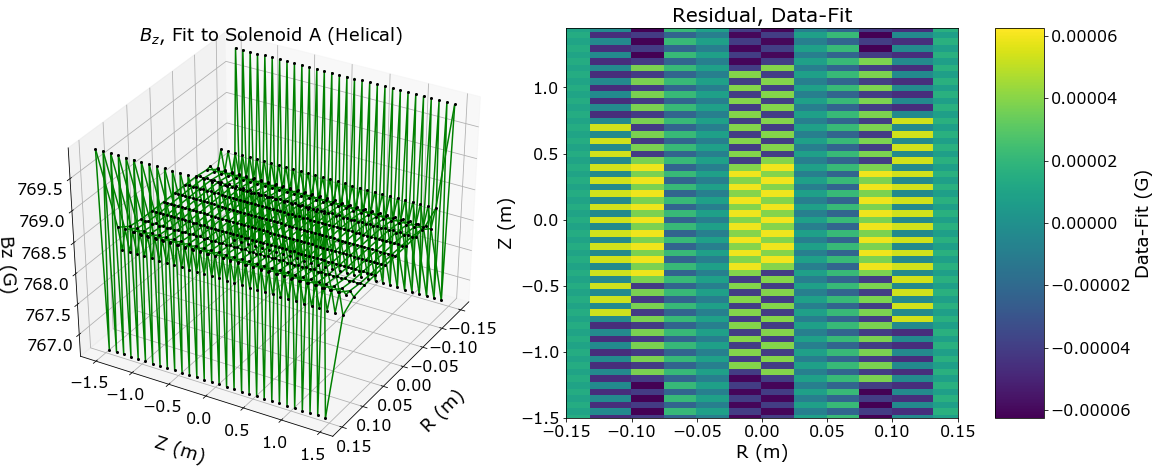}
    \includegraphics[width=\textwidth]{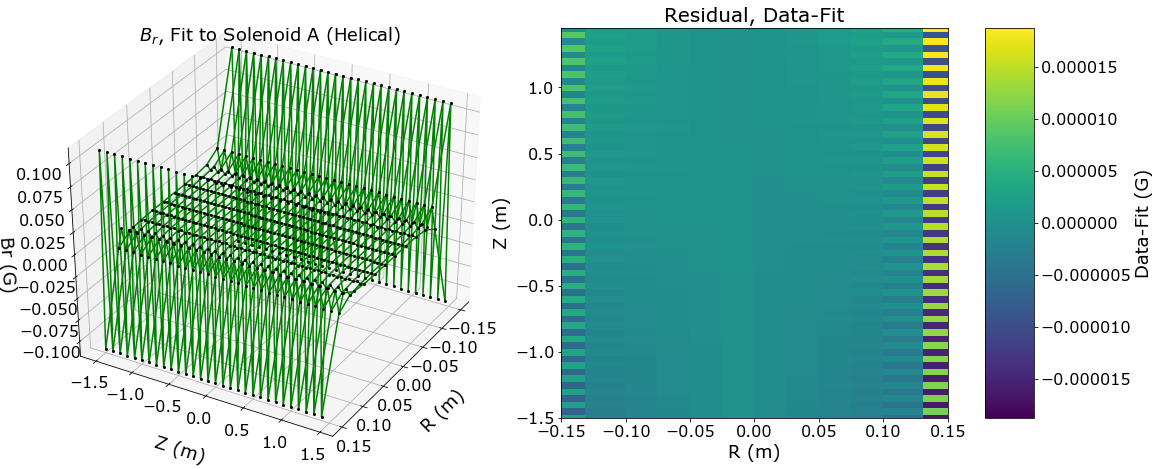}
    \includegraphics[width=\textwidth]{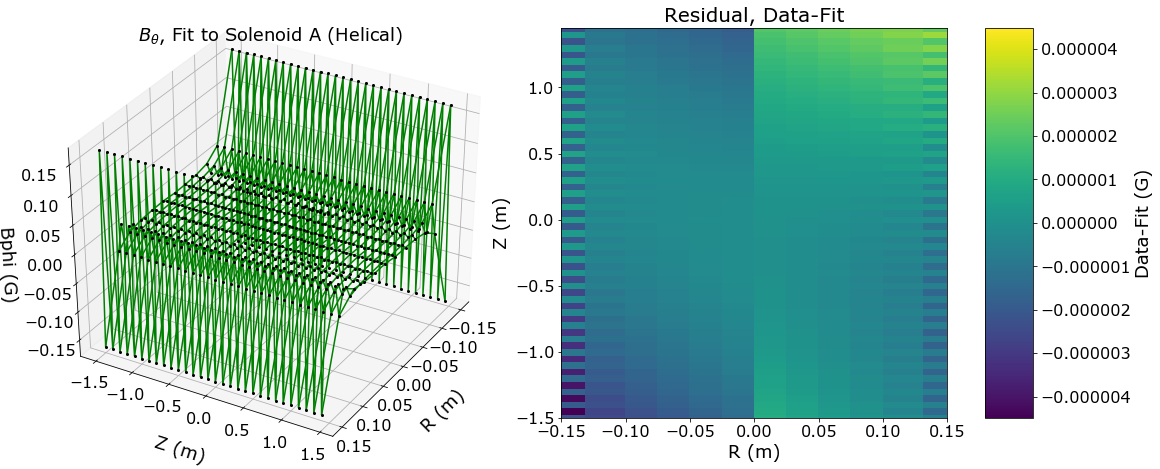}
    \centering
    \caption{Examples of fits using the helical harmonic function series to the magnetic field components, $B_z, B_r,
    B_{\theta}$, for 2D
    slices of solenoid A at a fixed angle $\theta=0$. The black points represent the data from the
    simulated solenoid, and the green mesh represents the fit. A residual is associated with each fit,
    showing the difference between data and fit in units of Gauss. (colour online)}\label{fig:fits_ex_A_hel}
\end{figure*}
\begin{figure*}
    \includegraphics[width=\textwidth]{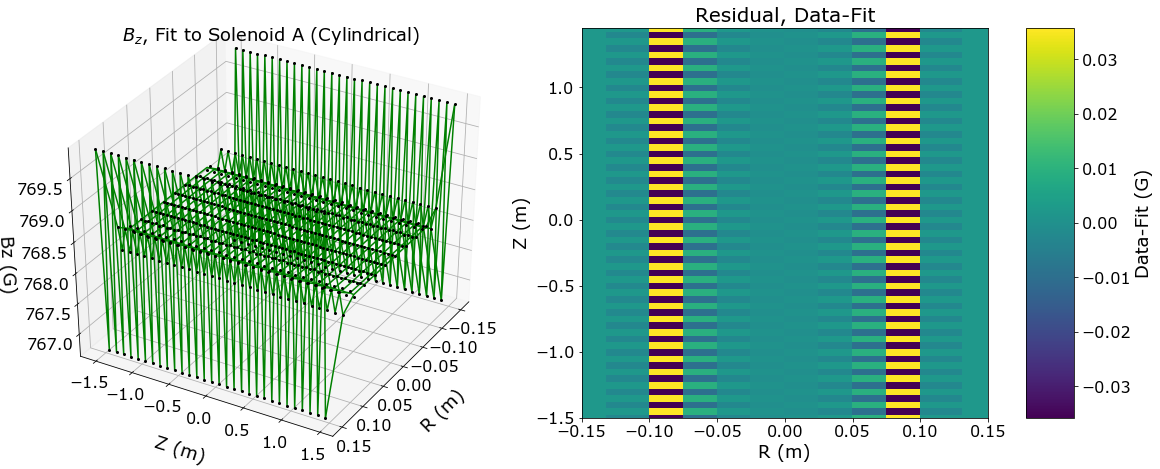}
    \includegraphics[width=\textwidth]{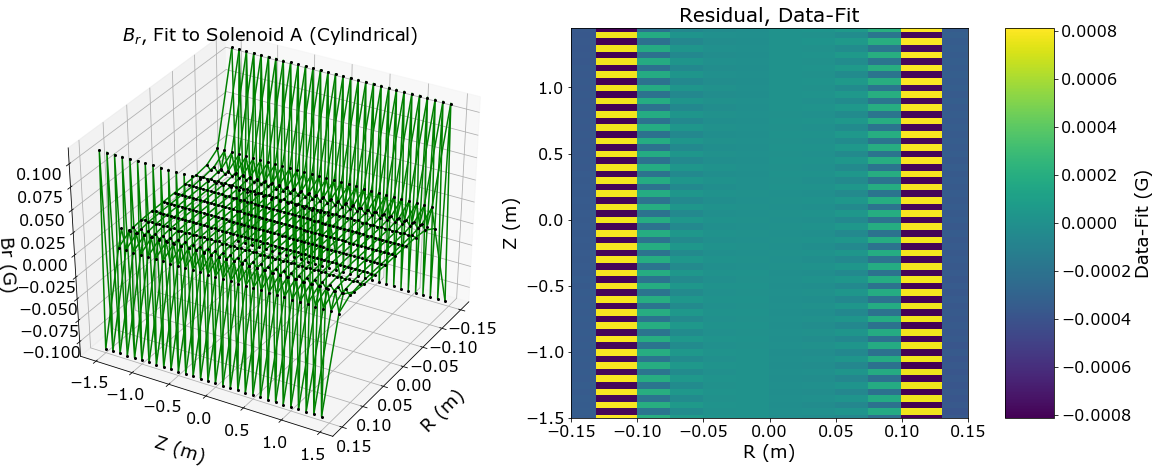}
    \includegraphics[width=\textwidth]{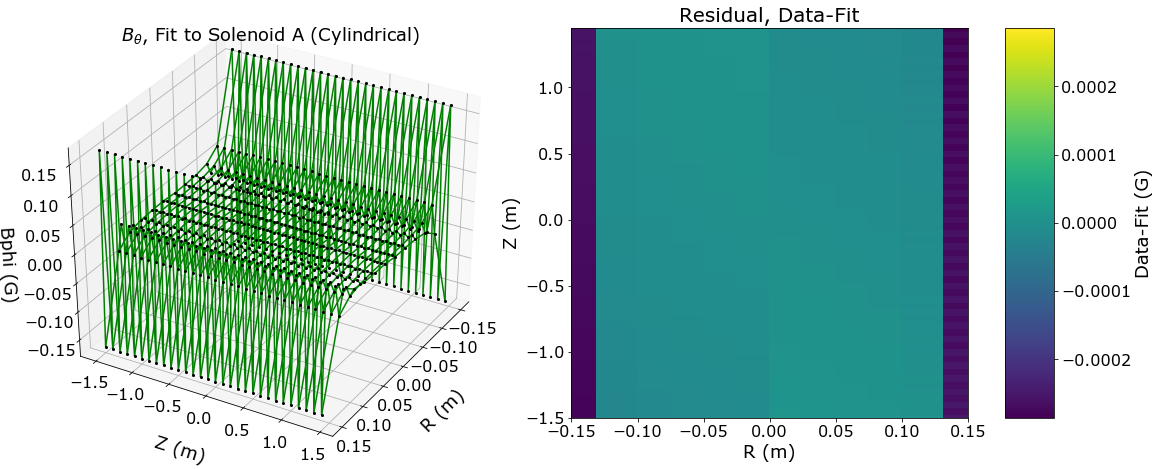}
    \centering
    \caption{Examples of fits using the cylindrical harmonic function series to the magnetic field components, $B_z, B_r,
    B_{\theta}$, for 2D
    slices of solenoid A at a fixed angle $\theta=0$. The black points represent the data from the
    simulated solenoid, and the green mesh represents the fit. A residual is associated with each fit,
    showing the difference between data and fit in units of Gauss. (colour online)}\label{fig:fits_ex_A_cyl}
\end{figure*}

An example of the fits, with associated residuals, are shown in
\cref{fig:fits_ex_A_hel,fig:fits_ex_A_cyl}.  The 3D plots on the left-hand side show the values of
the field components with both the calculated data (black points) and the fit function (green
mesh) for a single 2D plane of measurements.  Note that the parametric fit was determined using a
full 3D selection of points, not just the 2D plane displayed.  In all three components, the
wiggles due to the helical coil are significant, especially at the outer radius of the fit.  The size of the wiggles from
peak to trough are roughly 3-4\,G for the $B_z$ component and roughly 0.2-0.3\,G for the
$B_{\theta}$ and $B_r$ components.  The residuals, expressed as Data-Fit in units of Gauss, are
shown on the right-hand sides of the plots.  The vertical striping (especially prominent in the
upper right plot of \cref{fig:fits_ex_A_hel}, for example) corresponds to each of the radial
positions used in the sparse data set, as the residuals are only evaluated at each of the fitted
points for these plots.  The horizontal striping corresponds to the variation in the residuals as a
function of the $z$ step size.  While the left-hand plots for the helical and cylindrical fits may
look equivalent, the upper and lower bounds on the residuals for the helical fit are between two to
three orders of magnitude smaller than the those of the cylindrical fit.  The fit to helical harmonic functions is much more accurate than the fit to cylindrical functions.
\par
\begin{figure*}[!htb]
    \includegraphics[width=0.49\textwidth]{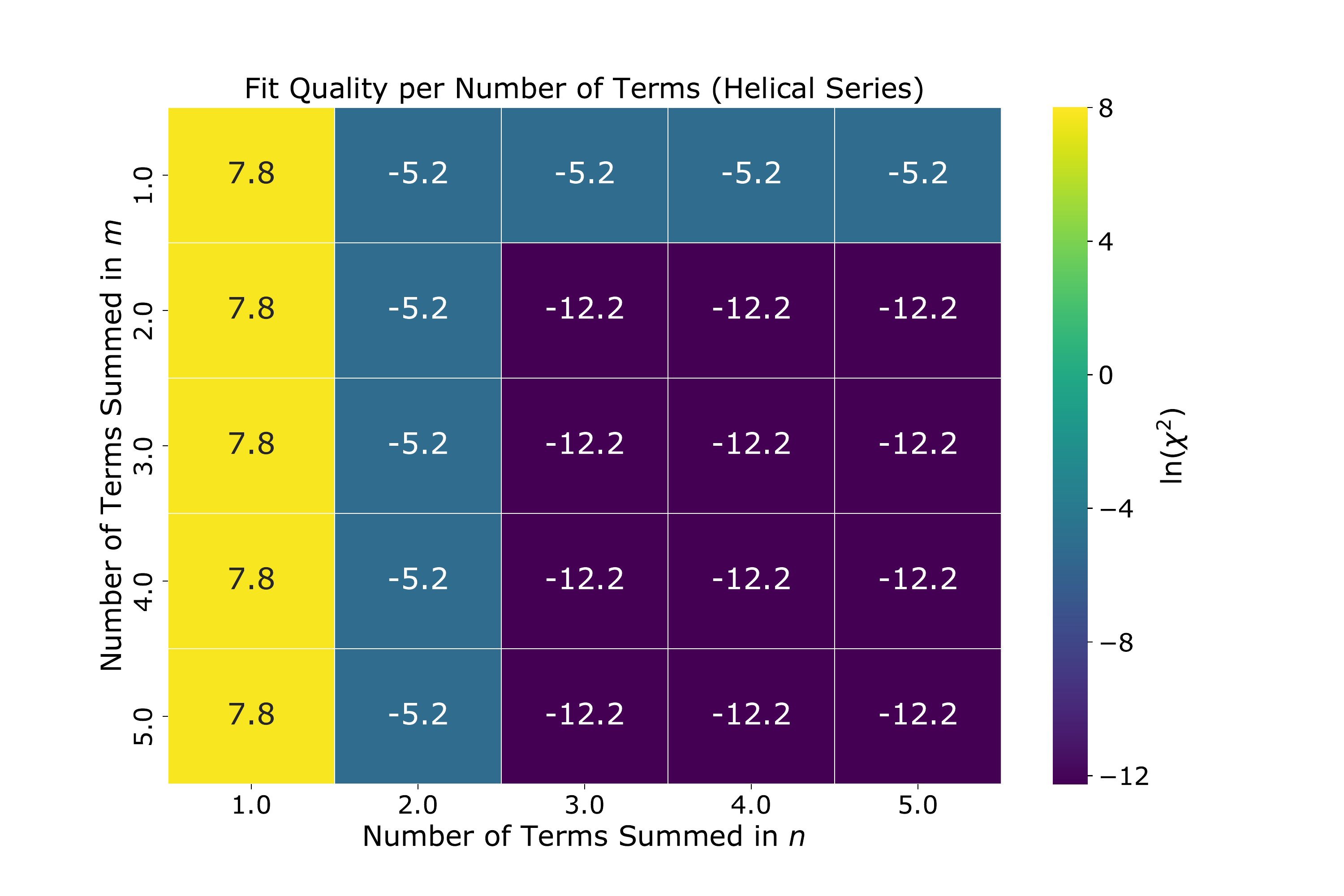}
    \includegraphics[width=0.49\textwidth]{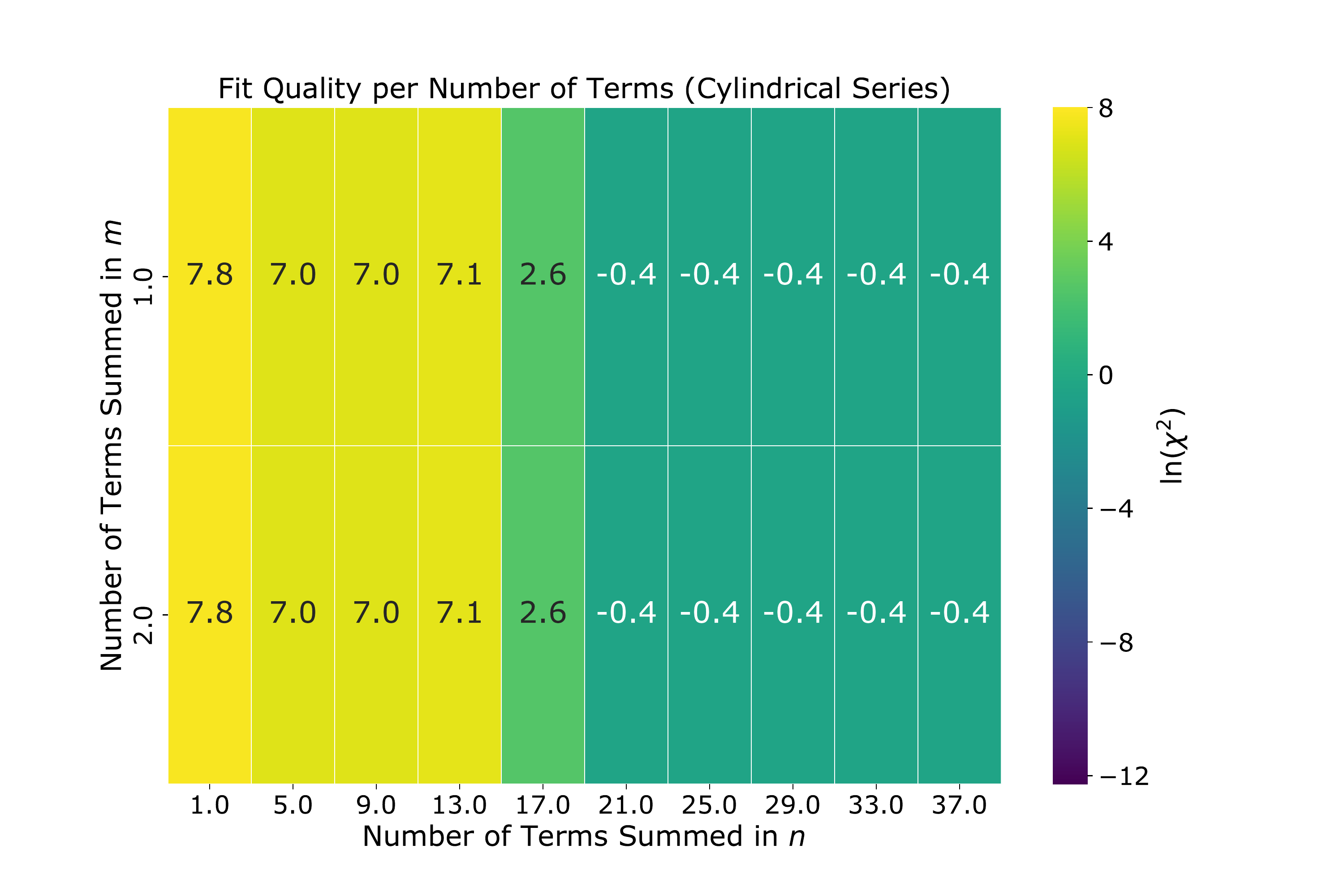}
    \centering
    \caption{Plots of $\ln(\chi^2)$ vs the number of $m$ and $n$ terms used in helical (left) and
        cylindrical (right) expansions for Solenoid A. A smaller $\ln(\chi^2)$ indicates a
        better quality of fit. The helical expansion requires far fewer parameters than the
        cylindrical expansion and obtains a superior quality of fit. (colour online)}\label{fig:chi2_solA}
\end{figure*}

The quality of the fits as a function of the total numbers of terms $m$ and $n$ is shown
in \cref{fig:chi2_solA}.  For both the cylindrical and helical expansions, the $m$ terms multiply
the $z$ coordinate inputs and the $n$ terms multiply the $\theta$ coordinate inputs.  The helical
series requires only a few terms in the expansion before it reaches a point where additional terms do not greatly improve the quality of fit.  In contrast, the cylindrical
series continually improves as the number of terms increases, until determining the optimal free
parameter values becomes too computationally taxing.  Even at roughly 10 times the number of free
parameters, the cylindrical expansion does not reach the level of fit quality found using the
helical expansion.
\par
\begin{figure*}[!htb]
    \includegraphics[width=0.49\textwidth]{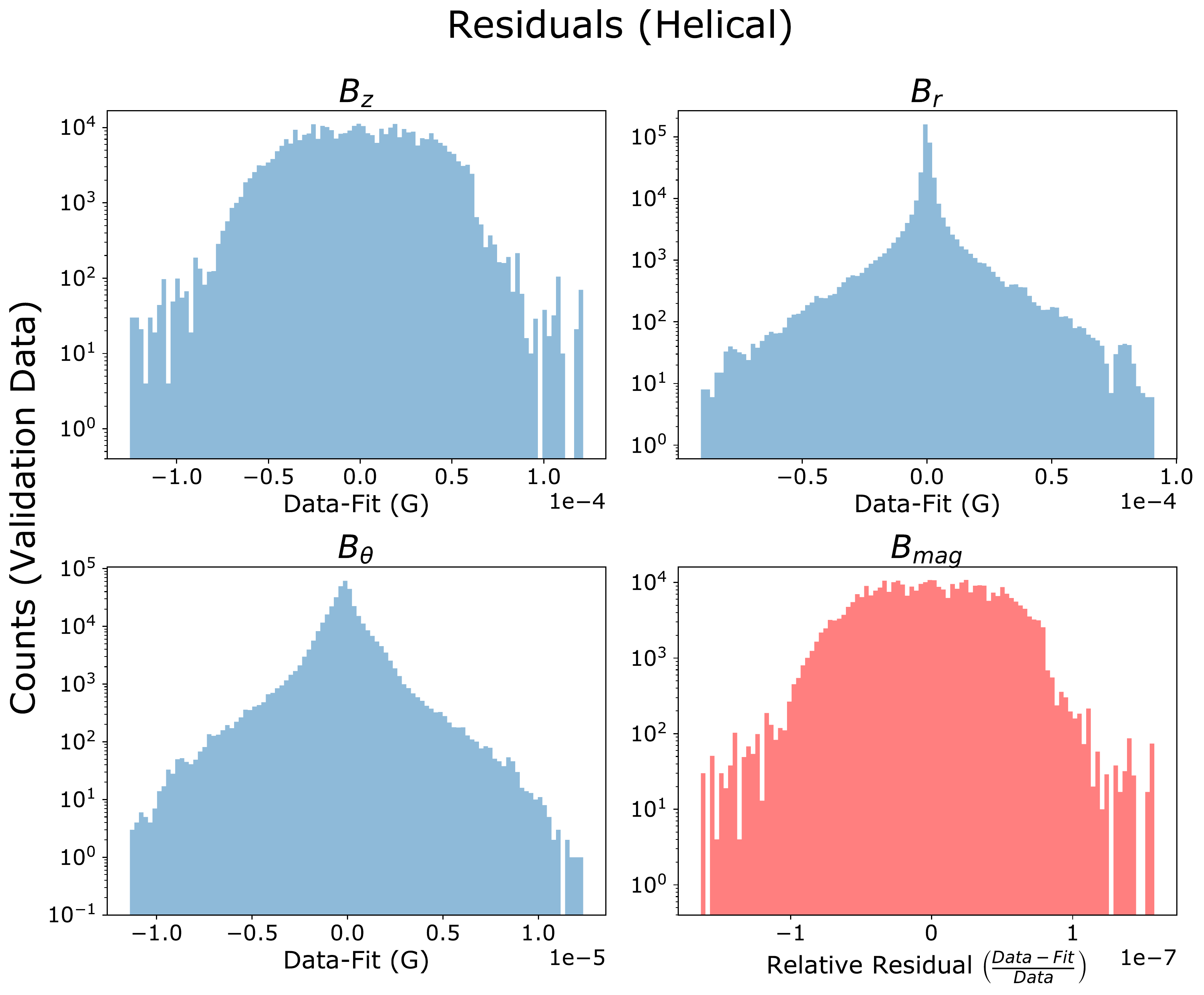}
    \includegraphics[width=0.49\textwidth]{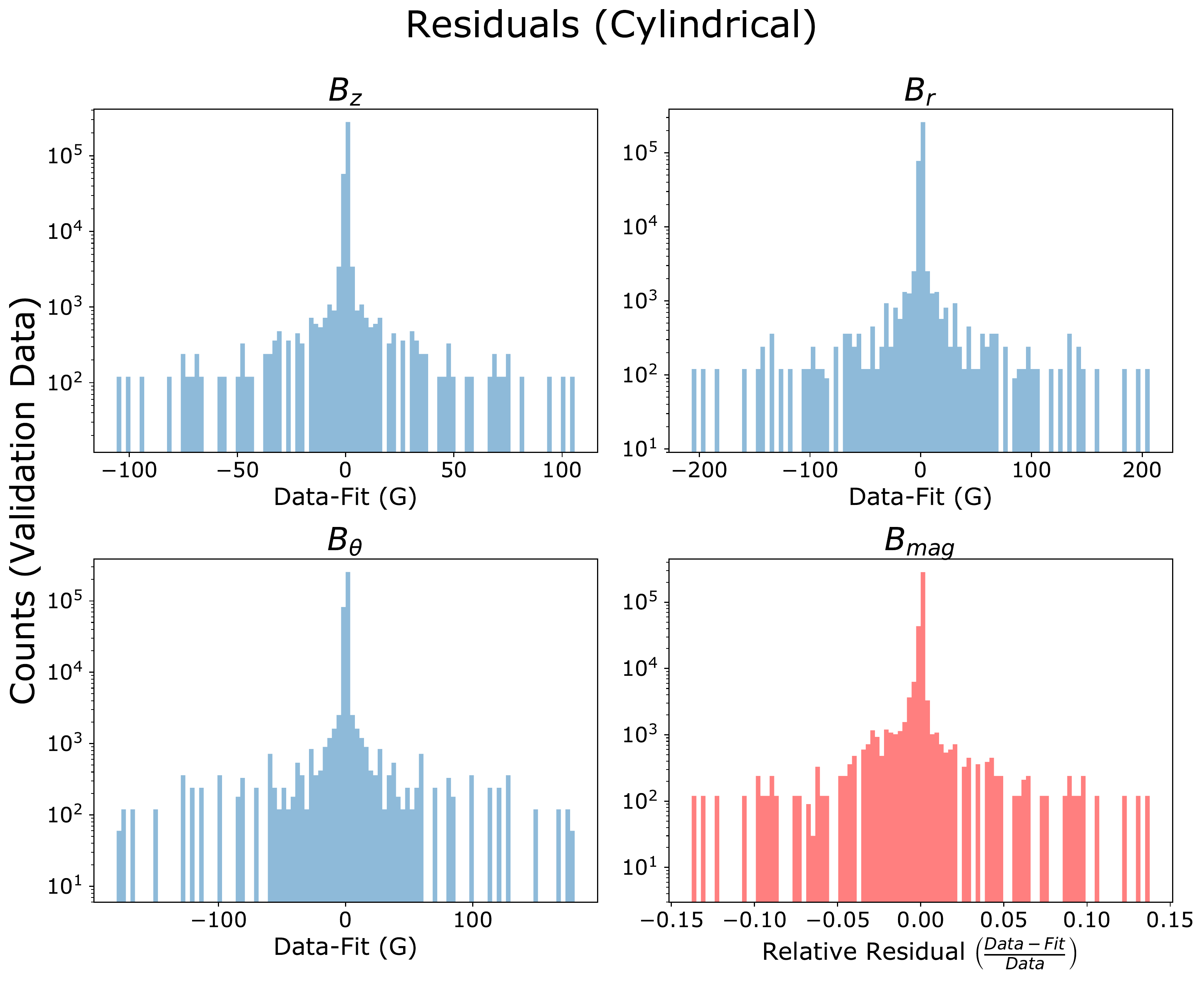}
    \centering
    \caption{Residuals for the individual field components and for the overall magnitude of the
        field, using the validation data set (Solenoid A).  Left: Helical Series. Right: Cylindrical
        series. Note that the horizontal axes for the cylindrical series cover a much larger range than the
        horizontal axes of the helical series. (colour online)}\label{fig:val_sol_A}
\end{figure*}

The residuals derived from the validation dataset are shown in
\cref{fig:val_sol_A} for both the helical and cylindrical series.  Note that the range of the
horizontal axis for the cylindrical and helical plots are vastly different; the range for the helical plot is
about six orders of magnitude smaller than the cylindrical. At its worst, the range of the residuals for the helical
series are at least four orders of magnitude better than the residuals for the cylindrical series, and
with roughly 10 times fewer free parameters.  As expected, the helical series is the superior choice
for this example, and the relative residuals on the overall magnitude of the field are typically
much better than 0.1 parts per million.
\par
The set of terms chosen to construct the validation plots (\cref{fig:val_sol_A}) was $m\leq 2$ and
$n\leq 3$.  Additional terms only added a one to two percent improvement to the $\chi^2$ statistic, as
can be seen from \cref{fig:chi2_solA}.
In \cref{subsec:lapace}, terms in which $n=m$ (or $n=m+1$, due to the redefinition in
\cref{sec:hel_func}) are shown to lead to solutions for constant helical surfaces for a given pitch
for a right handed helix.  The results in \cref{fig:chi2_solA} confirm that prediction; there is a
large improvement in the residuals for both the $(m,n) \leq (1,2)$ terms and the $(m,n) \leq (2,3)$
terms. The cylindrical series used for validation used $m\leq 1$ and $n\leq 21$, as additional terms
produced no improvement to the $\chi^2$ metric or the residual validation plots.

%\begin{figure*}[!htb]
%\begin{figure*}[!htb]

\FloatBarrier
\subsubsection{\label{sec:solB}Solenoid B}
\par
Solenoid B differs from Solenoid A in three distinct ways: it has a larger radius (1\,m), a
smaller pitch (7.5\,mm), and a shorter length (9.2\,m) than Solenoid A.  The field has a roughly 1\,T
$B_z$ component, and $B_r$ and $B_{\theta}$ components that are 3-5 orders of magnitude smaller.
These properties lead to a magnetic field which does not exhibit helically-induced wiggles
in the field except very close to the coil.  Additionally, the
large radius and small pitch lead to a large radius-to-pitch ratio, which presents
challenges to the helical harmonic series.  A very large ($\gg10$) radius-to-pitch ratio is difficult to use because it is the argument of the modified Bessel functions used in the
helical series expansion.  This large argument causes the Bessel functions to grow until they are
computationally unwieldy ($>10^{30}$).  The shorter length of the solenoid impacts the uniformity of
the field components as a function of $z$, such that they vary smoothly over the length of the
solenoid. In order to address all these features, the cylindrical series expansion is used instead
of the helical series, where the input length scale, $L$, is the total length of the solenoid.  The
results of this example are used to motivate the parameterization of the fit to Solenoid C, which is
a hybrid of Solenoids A and B.
\par
\begin{figure*}
    \includegraphics[width=\textwidth]{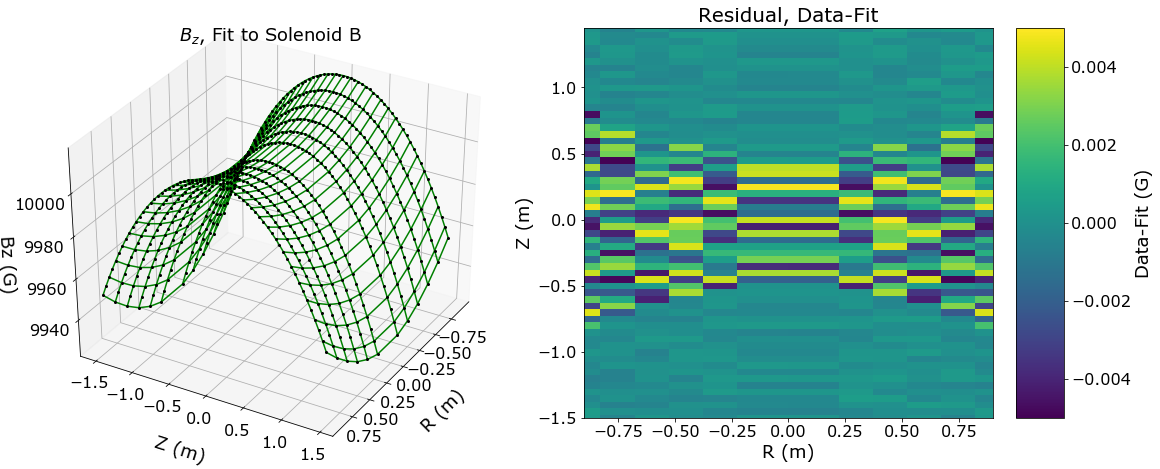}
    \includegraphics[width=\textwidth]{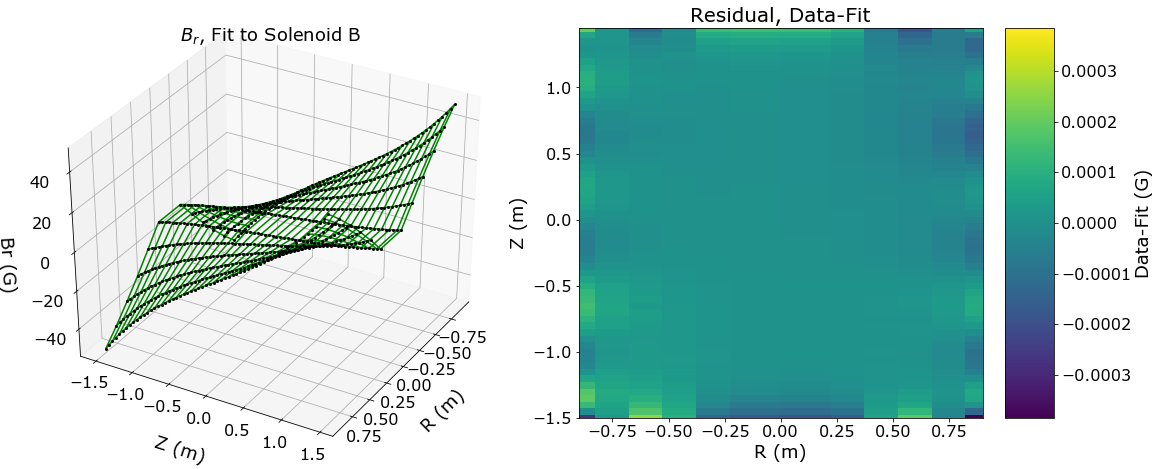}
    \includegraphics[width=\textwidth]{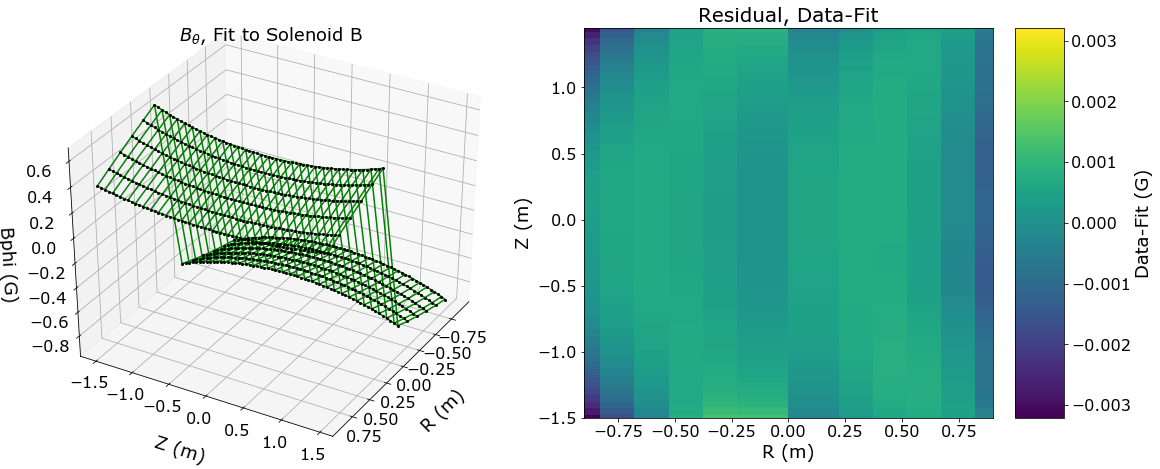}
    \centering
    \caption{Examples of fits using the cylindrical series to the magnetic field components, $B_z, B_r,
    B_{\theta}$, for 2D
    slices of solenoid A at a fixed angle $\theta$. The black points represent the data from the
    simulated solenoid, and the green mesh represents the fit. A residual is associated with each fit,
    showing the difference between data and fit in units of Gauss. (colour online)}\label{fig:fits_ex_B}
\end{figure*}
\begin{figure*}
    \includegraphics[width=0.75\textwidth]{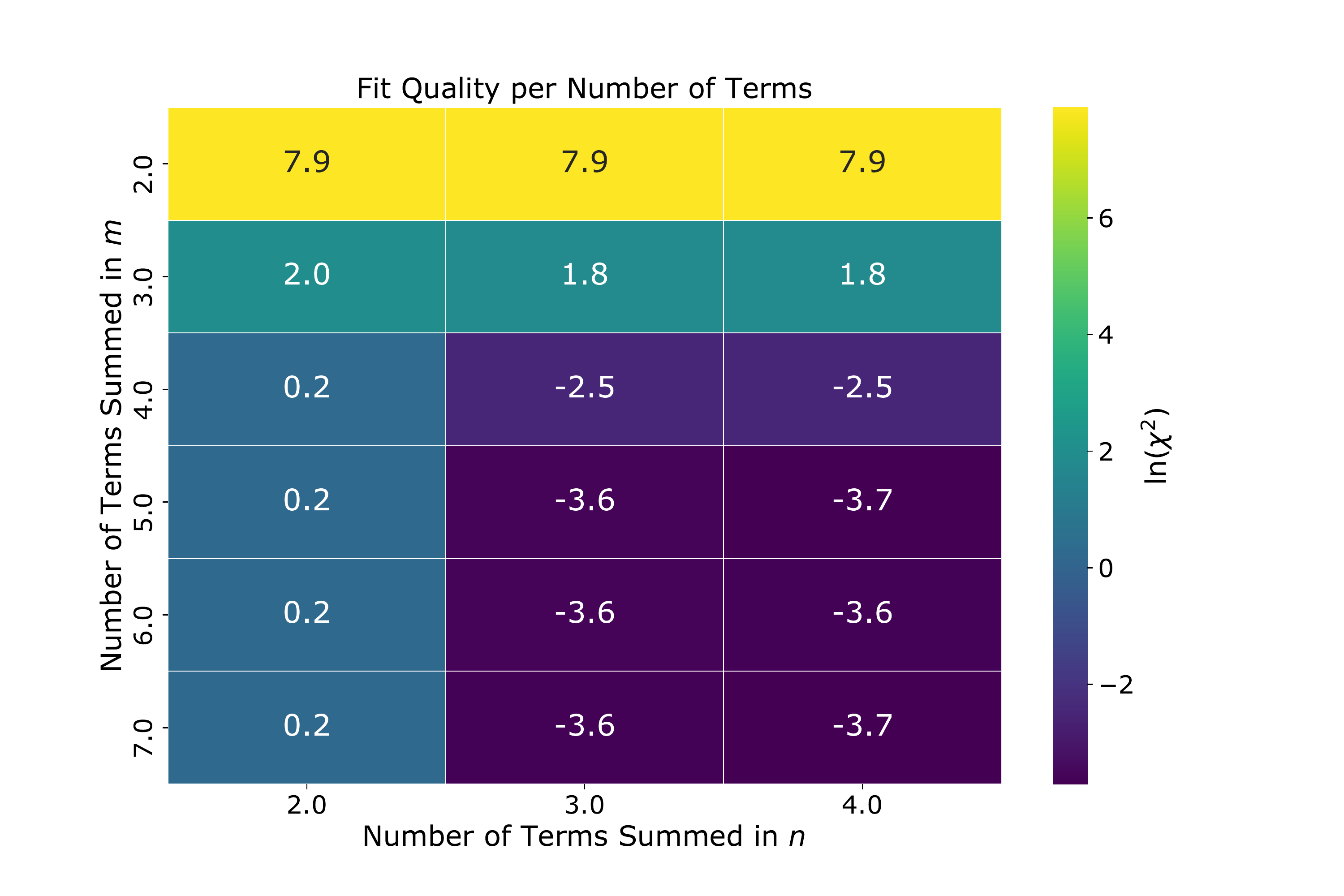}
    \centering
    \captionsetup{width=.75\textwidth}
    \caption{Plot of $\ln(\chi^2)$ vs the number of $m$ and $n$ terms used in 
        cylindrical expansions for Solenoid B. A smaller $\ln(\chi^2)$ indicates a
        better quality of fit. (colour online)}\label{fig:chi2_solB}
\end{figure*}

The sparse data set used for fitting consists of the same $\theta$ and $z$ selection used for
Solenoid A, along with 6 radial positions between 15 and 90\,cm, in 15\,cm increments. Examples of the
fits to Solenoid B are shown in \cref{fig:fits_ex_B}.  As for the to Solenoid A fits, the
horizontal and vertical striping in the residual plots is a consequence of the discrete nature of fit data.
The oscillations in the residuals, especially prominent in the $B_z$ residuals, are due to the
shorter wavelength $m$ terms used in the parameterization.  If more terms were added to the
parameterization, the magnitude of these oscillations would be reduced.  To determine the number of terms to include in the series expansion, the
fits were conducted iteratively, increasing the number of terms with each iteration.
The results of these fit iterations are shown in
\cref{fig:chi2_solB}.  Based on these results, the series is expanded to $m\leq5$ and $n\leq3$, as
additional terms only lead to diminishing (1-2\%) improvements to the residuals. 
\par
\begin{figure*}
    \includegraphics[width=0.75\textwidth]{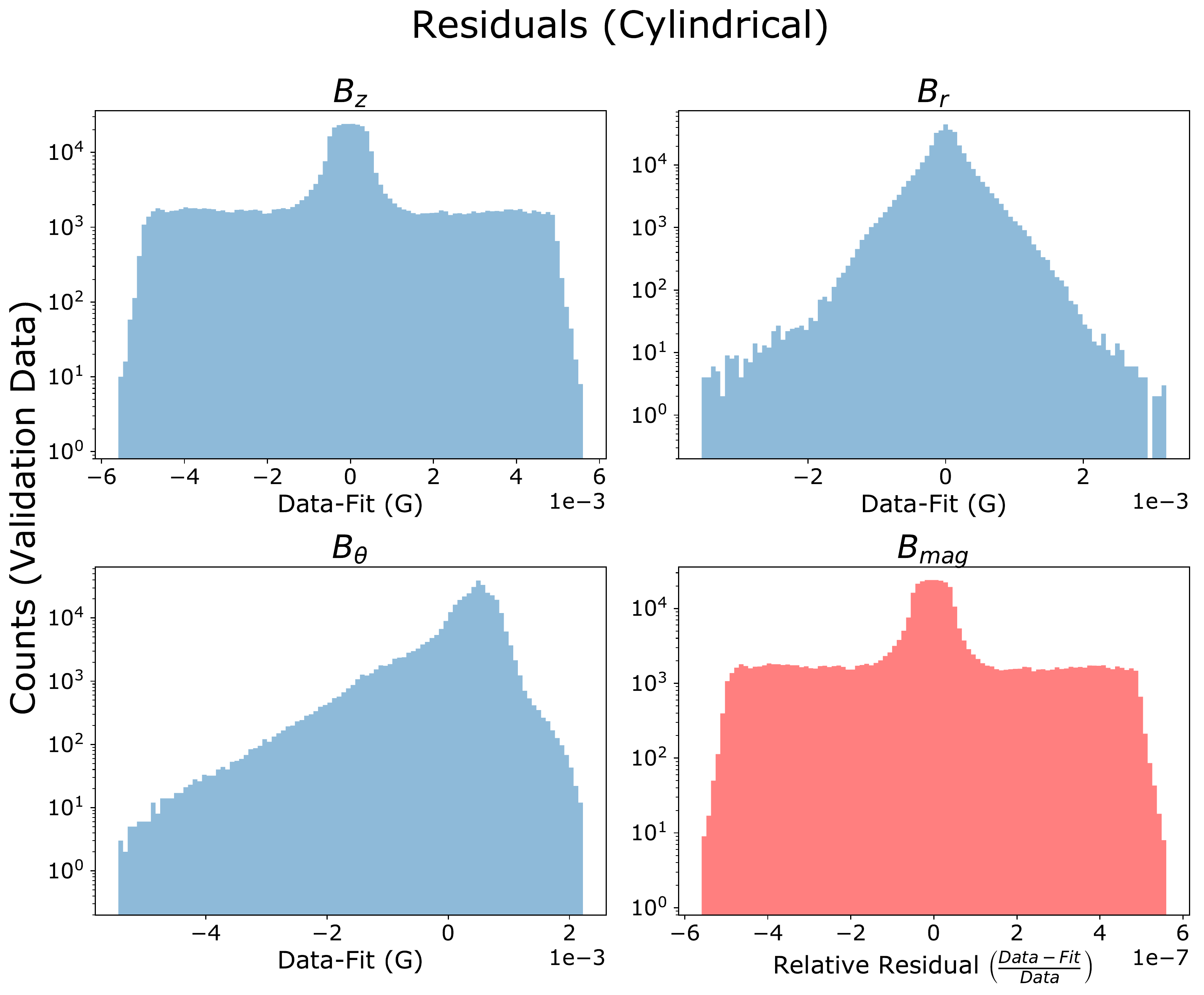}
    \centering
    \captionsetup{width=.75\textwidth}
    \captionof{figure}{Residuals for the individual field components and for the overall magnitude of the
        field, using the validation data set (Solenoid B). The skewness seen most prominently in $B_\theta$ is driven by the predictions at large $R$ and $|Z|$, see \cref{fig:fits_ex_B}. (colour online)}\label{fig:val_sol_B}

\end{figure*}

The validation set used to determine the per component residuals for this model is identical in
$\theta$ and $z$ with respect to Solenoid A, and used 24 equally spaced radial steps between 3.75\,cm
and 90\,cm.  The residuals based on the validation data are shown in \cref{fig:val_sol_B}.  The
residuals are never larger than $5\cdot10^{-3}$\,G, and are typically much smaller.  Overall, the
residuals are smaller than 1 part per million. The skewness
of the residual plot for $B_{\theta}$, which affects about 1\% of the validation dataset, is driven by an unavoidable asymmetry due to first and last turns of the solenoid, which deviate from an ideal cylinder. This can be seen in the $B_{\theta}$ residual plot of
\cref{fig:fits_ex_B}, specifically the darker blue regions in the upper and lower left corners
corresponding to the residuals at large radius and $|z|>1.0$\,m. 

%\begin{figure*}[!htb]

%\begin{figure*}[!htb]
\subsubsection{\label{sec:solC}Solenoid C}
\par
Solenoid C is identical to Solenoid A in all respects except for length;
it is the same length as Solenoid B, (10 times shorter than Solenoid A), which
enhances the $z$-dependent, $\theta$-independent features of the magnetic field due to the finite
nature of the solenoid. As can be seen in \cref{fig:mag_field_ex}, Solenoid C has both the large
scale, smooth variation of Solenoid B in addition to the high frequency wiggles of Solenoid A.  To model this simulation appropriately, a combination of terms is used from both the helical and
cylindrical series expansions.
\par
\begin{figure*}
    \includegraphics[width=\textwidth]{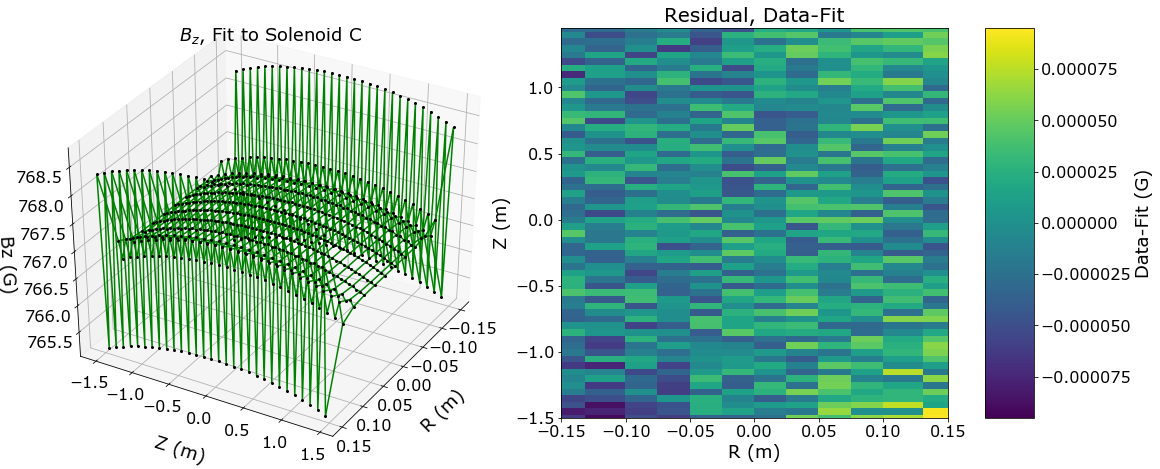}
    \includegraphics[width=\textwidth]{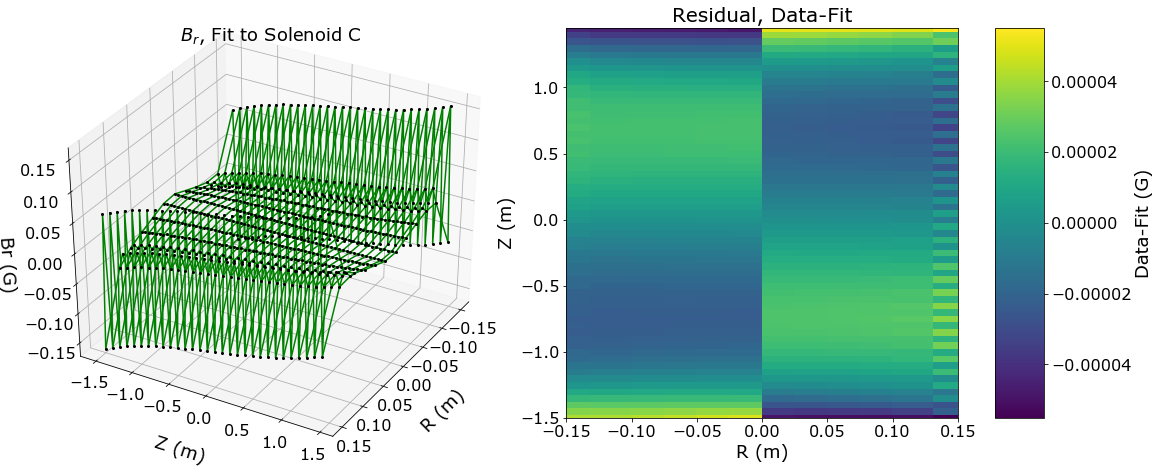}
    \includegraphics[width=\textwidth]{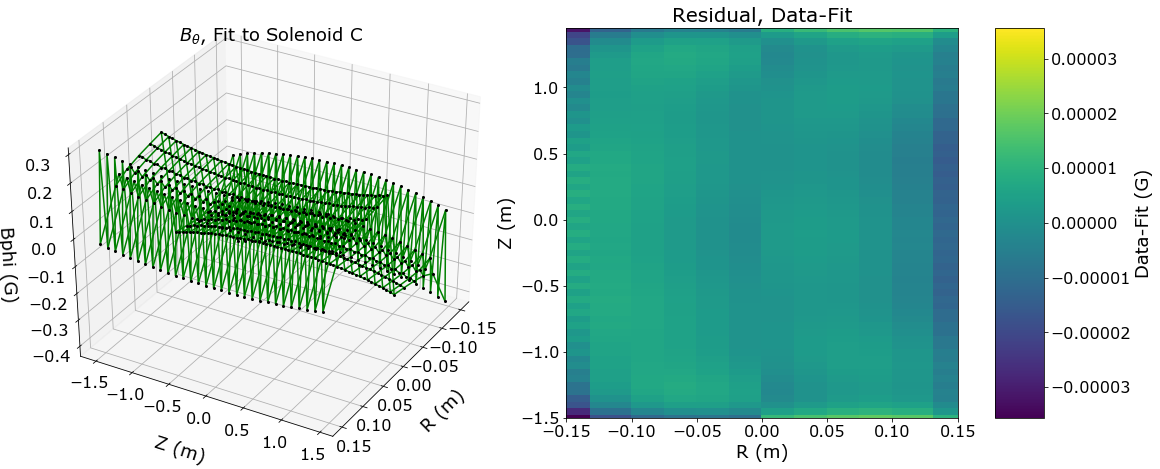}
    \centering
    \caption{Examples of fits using a combination of the helical and cylindrical series to the
        magnetic field components, $B_z, B_r, B_{\theta}$, for 2D slices of Solenoid C at a fixed
        angle $\theta$. The black points represent the data from the simulated solenoid, and the
        green mesh represents the fit. A residual is associated with each fit, showing the
    difference between data and fit in units of Gauss. (colour online)}\label{fig:fits_ex_C}
\end{figure*}
\begin{figure*}
    \includegraphics[width=0.65\textwidth]{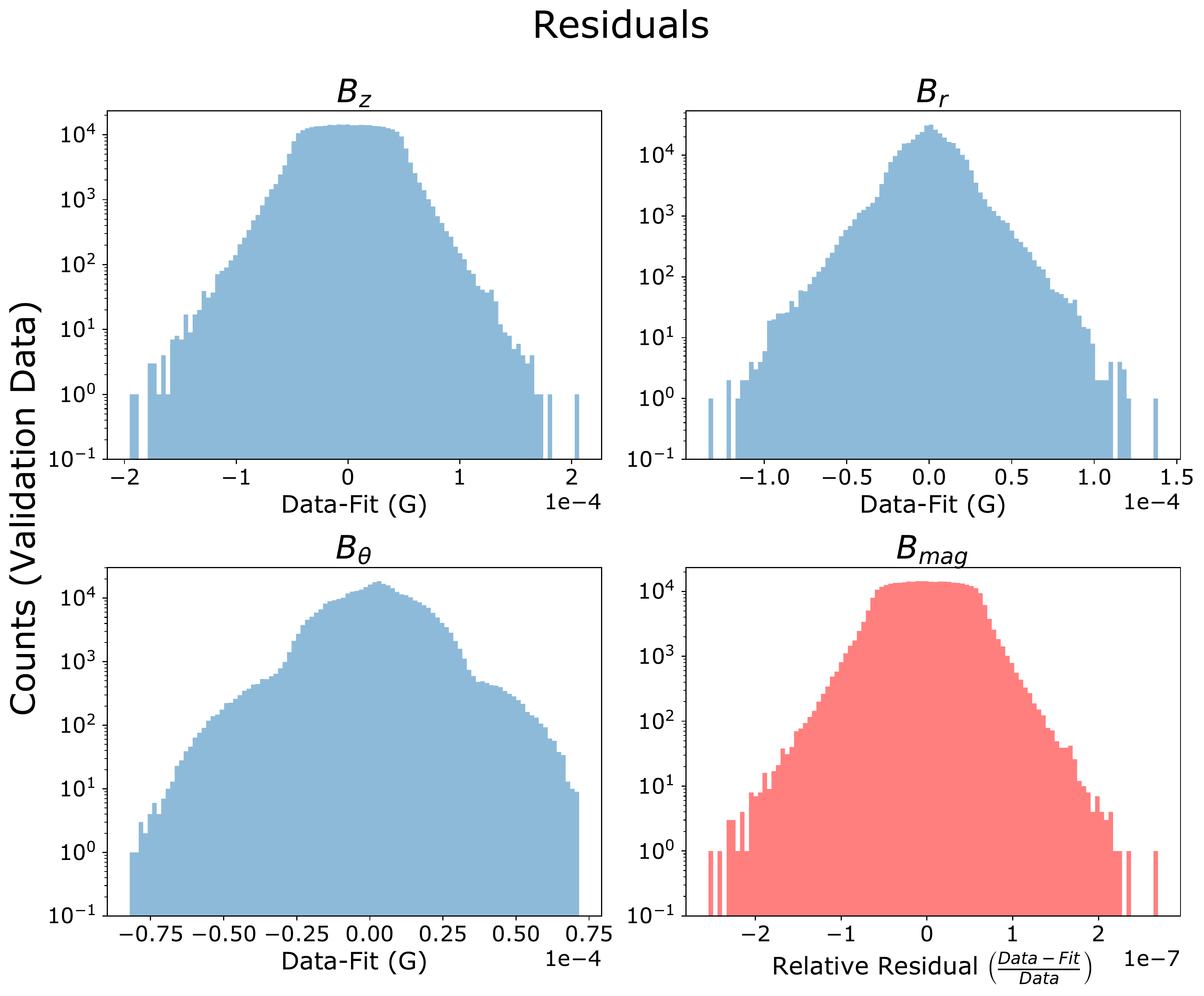}
    \centering
    \caption{Residuals for the individual field components and for the overall magnitude of the
        field, using the validation data set (Solenoid C). (colour online)}\label{fig:val_sol_C}
\end{figure*}

The sparse set of data selected to fit for Solenoid C, as well as the set of validation data, are
identical to the sets used for Solenoid A.  Instead of performing a scan over a very large potential
space of terms, the number of terms for Solenoid A ($m\leq2$ and $n\leq3$ for helical) is
combined with the number of terms for Solenoid B ($m\leq3$ and $n\leq5$ for cylindrical),
and all parameters are free.  Examples of the fit to Solenoid C are shown in
\cref{fig:fits_ex_C}.  The residual plots indicate that the fields are modeled very well, with only
small ($<10^{-4}$\,G) variations between fit and data.  The residuals based on the validation
dataset, shown in \cref{fig:val_sol_C}, are on the order of $10^{-4}$\,G or less for all field
components, and the residuals for the total field magnitude are better than one part per million,
which indicates that this combination of terms works well for modeling all major features of this
magnetic field.
\par
Two further studies were performed on Solenoid C. The first study investigated the effect of varying the density of the sparse data used for fitting, and the second study explores the effect of measurement error on the fit data. 

In the first study, the original sparse data for fitting Solenoid C is a cylindrical lattice whose coordinates are one of 6 radial positions, 16 angular positions, and 60 $z$ positions. The total number of data points is 5760. We can specify subgrids by selecting subsets of each of the radial, angular, and $z$ positions.

\begin{figure*}
    \includegraphics[width=0.65\textwidth]{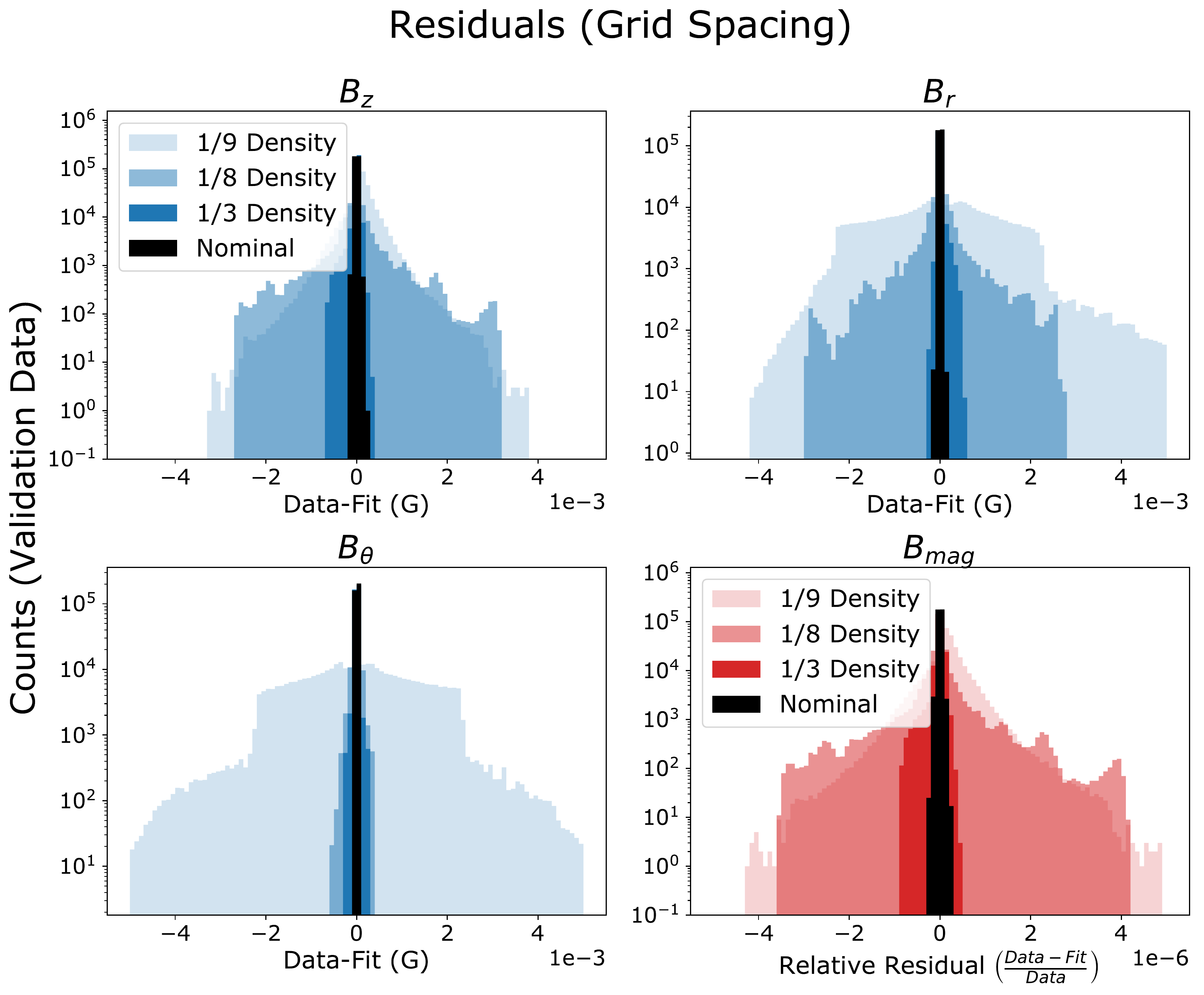}
    \centering
    \caption{Residuals for the individual field components and for the overall magnitude of the field, using the validation data set (Solenoid C). Each subplot has the results of the three simulations which vary the fitting grid density, darker shades correspond to a denser grid. The residuals for the original nominal fitting grid are overlaid in black. (colour online)}\label{fig:grid_dens_sol_C}
\end{figure*}

\cref{fig:grid_dens_sol_C} shows the residuals for fits using subgrids with 1/3, 1/8, and 1/9 of the density of the original fitting grid. The 1/3 and 1/9 density subgrids are formed by selecting every $3^{rd}$ and every $9^{th}$ point among the $z$ positions, respectively. The 1/8 density subgrid is formed by taking every other point in each of the radial, angular, and $z$ positions. It is clear that the error increases as the density decreases, but even the largest relative error in magnitude is still on the order $10^{-6}$ for the sparsest grid. We also note that the original grid is already a very coarse subgrid of the fine grid used for validation.  The details of these grids are shown in \cref{tab:grids}.

\begin{tabular}{|c|c|c|}
\hline
 Grid & Density & Percent of total \\\hline
Validation grid & ~~~362,880 points & ~~~100\% \\\hline
Nominal fitting grid & 5760 points & ~~1.587\%    \\\hline
1/3 density grid & 1920 points & 0.529\%     \\\hline
1/8 density grid & 720 points & 0.198\%      \\\hline
1/9 density grid & 640 points & 0.176\% \\\hline
\end{tabular}
\label{tab:grids}

\par
The second study examines the response of our method to measurement error on the fit data. 
For modern solenoid mapping, NMR and Hall probes are required to have measurement accuracy on the order of $10^{-5}$~\cite{hall_probe}. In order to simulate this type of measurement error, Gaussian noise is added independently to each field component in the data before fitting. The Gaussian mean is 0 and the standard deviation in each component is chosen to simulate a relative error of $\varepsilon$.  The residuals are calculated by comparing the fit values to the true field values ($\varepsilon=0$).

\begin{figure*}
    \includegraphics[width=0.65\textwidth]{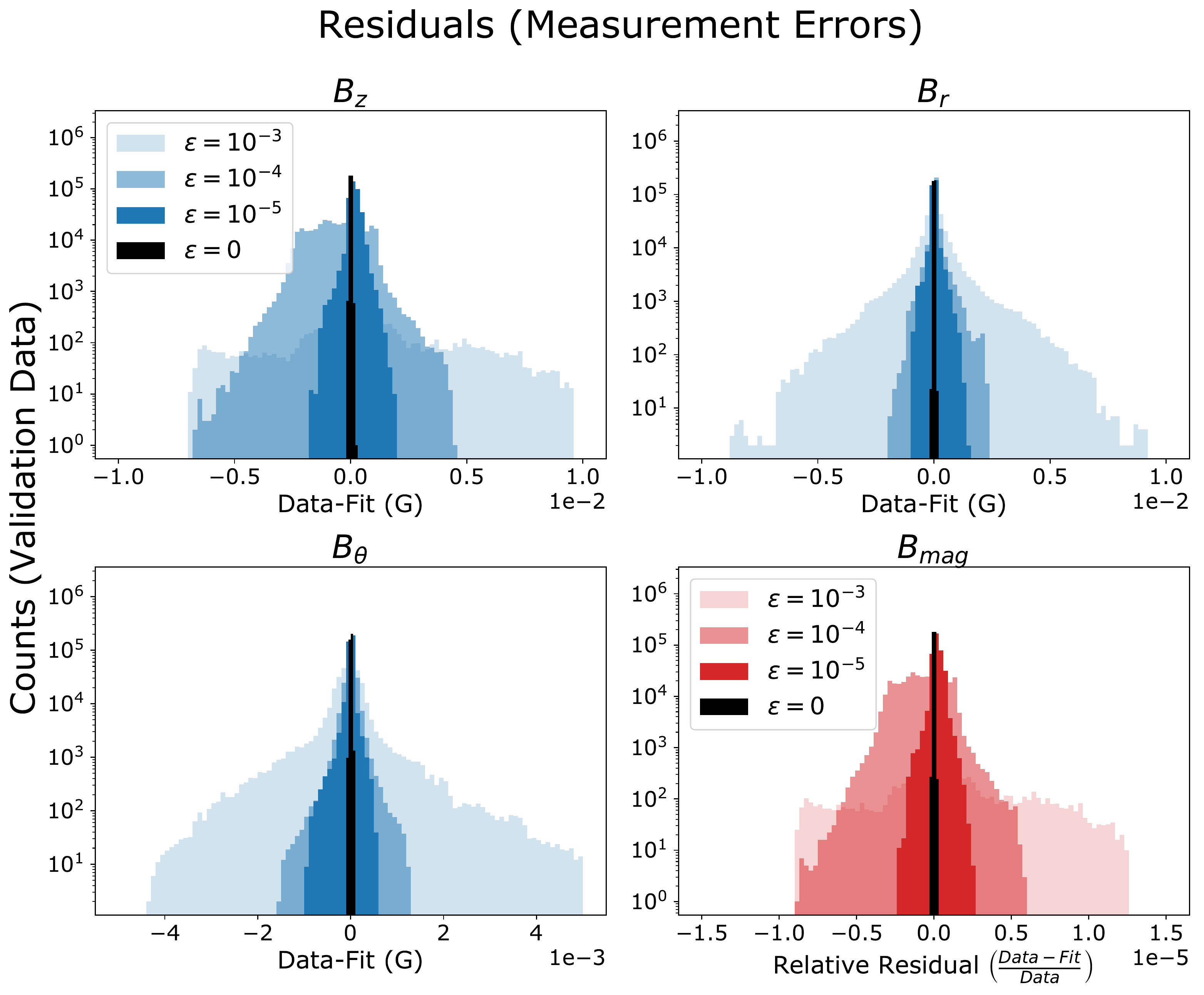}
    \centering
    \caption{Residuals for the individual field components and for the overall magnitude of the field, using the validation data set (Solenoid C). Each subplot has the results of the three simulations which vary the measurement error, darker shades correspond to less error. The residuals for the original measurements are overlaid in black. (colour online)}\label{fig:syst_err_sol_C}
\end{figure*}

\cref{fig:syst_err_sol_C} shows histograms of the residuals for fits with $\varepsilon = 10^{-3}, 10^{-4}, 10^{-5}, 0$. The histogram for $\varepsilon=0$ corresponds to the nominal fit as seen in \cref{fig:val_sol_C}.  The $\varepsilon = 10^{-5}$ error leads to a relative residual of better than $10^{-6}$ for the majority of data points, and even the most pessimistic error simulations are typically less than $10^{-5}$.  This implies that our method is robust with respect to expected measurement errors for magnetic field surveys.

    \section{\label{sec:conclusion}Summary and Conclusions}
\par
Starting from first principles, we have introduced general solutions to Laplace's equation in a helical coordinate system for the purposes of modeling realistic solenoidal magnetic fields.
Using sparse sets of data generated for different solenoidal geometries, we modeled the magnetic
field using expansions of helical and cylindrical harmonic functions and validated the model with a high-granularity magnetic field calculation.
Using a small number of free parameters, we were able to obtain accurate
magnetic field models, and capture asymmetric behavior caused by the helical coil.  The residuals are better than one part per million, and typically better than 0.1 parts per million.
\par
The helical series expansion is good at capturing the wiggles that arise from the helical winding of a solenoid.  Depending on the pitch and radius of the
solenoid, these oscillations can become significant compared to the average field
strength, and ignoring these contributions would lead to an incorrect model of the magnetic
field.  The cylindrical series are useful for modeling the low frequency components of a
finite-length solenoid.  Using an effective length equal to the solenoid length, the
cylindrical series models the $z$-dependent,
axially symmetric features of the magnetic field.  For solenoids of both finite length and with helical features, a combination of helical and cylindrical series solutions provides
the most accurate model.  While not exhaustive, we examined the impact of realistic measurement errors on our model.  Even with pessimistic error assumptions, our model still performs well, with residuals better than 10 parts per million, and typically better than 1 part per million.  Further studies of magnetic field perturbation due to machining tolerances, external magnetic field sources (e.g. due to additional circuitry), and other sources of error are left for future investigation.
\par
While this method was shown to be useful for representing a magnetic field, it can
also be used to calculate a magnetic field.  The calculated fields used in this paper were generated using a
common numerical integration of the Biot-Savart law.  This process, even when
parallelized across GPUs, is computationally taxing, and its resource usage scales both as
a function of the size of the solenoid and the number of grid points in which one would like to
evaluate the field.  Conversely, starting with a closed-form solution of a complex magnetic field
scales to the number of grid points, and the evaluation of each grid point requires executing
only a handful of mathematical terms instead of a series of numeric integrals.  While the method for
determining the coefficients of this series without the use of fitting is not explored in
this paper, it should be possible for an appropriately specified field calculation to do so.
%\par
%There are many future studies in practice for further investigation and implementation of the
%mathematical 

    \\
    \begin{acknowledgments}
        The authors would like to thank Matt Bonakdarpour for helpful discussions. We gratefully
        acknowledge the support provided by the Department of Energy under award number
        DE-SC0015910. This document was prepared by members of the Mu2e Collaboration using the
        resources of the Fermi National Accelerator Laboratory (Fermilab), a U.S. Department of
        Energy, Office of Science, HEP User Facility. Fermilab is managed by Fermi Research
        Alliance, LLC (FRA), acting under Contract No. DE-AC02-07CH11359.  We thank the reviewers for  providing helpful comments on earlier drafts of the manuscript.
    \end{acknowledgments}

    \bibliographystyle{apsrev4-1}
    \bibliography{main}% Produces the bibliography via BibTeX.

\begin{thebibliography}{17}%
\makeatletter
\providecommand \@ifxundefined [1]{%
 \@ifx{#1\undefined}
}%
\providecommand \@ifnum [1]{%
 \ifnum #1\expandafter \@firstoftwo
 \else \expandafter \@secondoftwo
 \fi
}%
\providecommand \@ifx [1]{%
 \ifx #1\expandafter \@firstoftwo
 \else \expandafter \@secondoftwo
 \fi
}%
\providecommand \natexlab [1]{#1}%
\providecommand \enquote  [1]{``#1''}%
\providecommand \bibnamefont  [1]{#1}%
\providecommand \bibfnamefont [1]{#1}%
\providecommand \citenamefont [1]{#1}%
\providecommand \href@noop [0]{\@secondoftwo}%
\providecommand \href [0]{\begingroup \@sanitize@url \@href}%
\providecommand \@href[1]{\@@startlink{#1}\@@href}%
\providecommand \@@href[1]{\endgroup#1\@@endlink}%
\providecommand \@sanitize@url [0]{\catcode `\\12\catcode `\$12\catcode
  `\&12\catcode `\#12\catcode `\^12\catcode `\_12\catcode `\%12\relax}%
\providecommand \@@startlink[1]{}%
\providecommand \@@endlink[0]{}%
\providecommand \url  [0]{\begingroup\@sanitize@url \@url }%
\providecommand \@url [1]{\endgroup\@href {#1}{\urlprefix }}%
\providecommand \urlprefix  [0]{URL }%
\providecommand \Eprint [0]{\href }%
\providecommand \doibase [0]{http://dx.doi.org/}%
\providecommand \selectlanguage [0]{\@gobble}%
\providecommand \bibinfo  [0]{\@secondoftwo}%
\providecommand \bibfield  [0]{\@secondoftwo}%
\providecommand \translation [1]{[#1]}%
\providecommand \BibitemOpen [0]{}%
\providecommand \bibitemStop [0]{}%
\providecommand \bibitemNoStop [0]{.\EOS\space}%
\providecommand \EOS [0]{\spacefactor3000\relax}%
\providecommand \BibitemShut  [1]{\csname bibitem#1\endcsname}%
\let\auto@bib@innerbib\@empty
%</preamble>
\bibitem [{\citenamefont {Feher}\ \emph {et~al.}(2018)\citenamefont {Feher},
  \citenamefont {DeLurgio}, \citenamefont {Elementi}, \citenamefont {Friedsam},
  \citenamefont {Grudzinski}, \citenamefont {Lamm}, \citenamefont {Nogiec},
  \citenamefont {Orozco}, \citenamefont {Pollack}, \citenamefont {Schmitt},
  \citenamefont {Strauss}, \citenamefont {Talaga}, \citenamefont {Wagner},
  \citenamefont {White},\ and\ \citenamefont {Zhao}}]{mu2e}%
  \BibitemOpen
  \bibfield  {author} {\bibinfo {author} {\bibfnamefont {S.}~\bibnamefont
  {Feher}}, \bibinfo {author} {\bibfnamefont {P.}~\bibnamefont {DeLurgio}},
  \bibinfo {author} {\bibfnamefont {L.}~\bibnamefont {Elementi}}, \bibinfo
  {author} {\bibfnamefont {H.~W.}\ \bibnamefont {Friedsam}}, \bibinfo {author}
  {\bibfnamefont {J.~J.}\ \bibnamefont {Grudzinski}}, \bibinfo {author}
  {\bibfnamefont {M.~J.}\ \bibnamefont {Lamm}}, \bibinfo {author}
  {\bibfnamefont {J.~M.}\ \bibnamefont {Nogiec}}, \bibinfo {author}
  {\bibfnamefont {C.}~\bibnamefont {Orozco}}, \bibinfo {author} {\bibfnamefont
  {B.}~\bibnamefont {Pollack}}, \bibinfo {author} {\bibfnamefont {M.~H.}\
  \bibnamefont {Schmitt}}, \bibinfo {author} {\bibfnamefont {T.}~\bibnamefont
  {Strauss}}, \bibinfo {author} {\bibfnamefont {R.~L.}\ \bibnamefont {Talaga}},
  \bibinfo {author} {\bibfnamefont {R.~G.}\ \bibnamefont {Wagner}}, \bibinfo
  {author} {\bibfnamefont {J.~L.}\ \bibnamefont {White}}, \ and\ \bibinfo
  {author} {\bibfnamefont {H.}~\bibnamefont {Zhao}},\ }\href {\doibase
  10.1109/TASC.2017.2786720} {\bibfield  {journal} {\bibinfo  {journal} {IEEE
  Transactions on Applied Superconductivity}\ }\textbf {\bibinfo {volume}
  {28}},\ \bibinfo {pages} {1} (\bibinfo {year} {2018})}\BibitemShut {NoStop}%
\bibitem [{\citenamefont {Nouri}\ and\ \citenamefont
  {Plaster}(2013)}]{NOURI201330}%
  \BibitemOpen
  \bibfield  {author} {\bibinfo {author} {\bibfnamefont {N.}~\bibnamefont
  {Nouri}}\ and\ \bibinfo {author} {\bibfnamefont {B.}~\bibnamefont
  {Plaster}},\ }\href {\doibase https://doi.org/10.1016/j.nima.2013.05.013}
  {\bibfield  {journal} {\bibinfo  {journal} {Nuclear Instruments and Methods
  in Physics Research Section A: Accelerators, Spectrometers, Detectors and
  Associated Equipment}\ }\textbf {\bibinfo {volume} {723}},\ \bibinfo {pages}
  {30 } (\bibinfo {year} {2013})}\BibitemShut {NoStop}%
\bibitem [{\citenamefont {Gu}\ \emph {et~al.}(2011)\citenamefont {Gu},
  \citenamefont {Okamura}, \citenamefont {Pikin}, \citenamefont {Fischer},\
  and\ \citenamefont {Luo}}]{GU2011190}%
  \BibitemOpen
  \bibfield  {author} {\bibinfo {author} {\bibfnamefont {X.}~\bibnamefont
  {Gu}}, \bibinfo {author} {\bibfnamefont {M.}~\bibnamefont {Okamura}},
  \bibinfo {author} {\bibfnamefont {A.}~\bibnamefont {Pikin}}, \bibinfo
  {author} {\bibfnamefont {W.}~\bibnamefont {Fischer}}, \ and\ \bibinfo
  {author} {\bibfnamefont {Y.}~\bibnamefont {Luo}},\ }\href {\doibase
  https://doi.org/10.1016/j.nima.2011.01.123} {\bibfield  {journal} {\bibinfo
  {journal} {Nuclear Instruments and Methods in Physics Research Section A:
  Accelerators, Spectrometers, Detectors and Associated Equipment}\ }\textbf
  {\bibinfo {volume} {637}},\ \bibinfo {pages} {190 } (\bibinfo {year}
  {2011})}\BibitemShut {NoStop}%
\bibitem [{\citenamefont {Wittgenstein}\ \emph {et~al.}(1990)\citenamefont
  {Wittgenstein}, \citenamefont {Herve}, \citenamefont {Feldmann},
  \citenamefont {Luckey},\ and\ \citenamefont {Vetlitsky}}]{L3}%
  \BibitemOpen
  \bibfield  {author} {\bibinfo {author} {\bibfnamefont {F.}~\bibnamefont
  {Wittgenstein}}, \bibinfo {author} {\bibfnamefont {A.}~\bibnamefont {Herve}},
  \bibinfo {author} {\bibfnamefont {M.}~\bibnamefont {Feldmann}}, \bibinfo
  {author} {\bibfnamefont {D.}~\bibnamefont {Luckey}}, \ and\ \bibinfo {author}
  {\bibfnamefont {I.}~\bibnamefont {Vetlitsky}},\ }\enquote {\bibinfo {title}
  {Construction of the l3 magnet},}\ in\ \href {\doibase
  10.1007/978-94-009-0769-0_22} {\emph {\bibinfo {booktitle} {11th
  International Conference on Magnet Technology (MT-11): Volume 1}}},\ \bibinfo
  {editor} {edited by\ \bibinfo {editor} {\bibfnamefont {T.}~\bibnamefont
  {Sekiguchi}}\ and\ \bibinfo {editor} {\bibfnamefont {S.}~\bibnamefont
  {Shimamoto}}}\ (\bibinfo  {publisher} {Springer Netherlands},\ \bibinfo
  {address} {Dordrecht},\ \bibinfo {year} {1990})\ pp.\ \bibinfo {pages}
  {130--135}\BibitemShut {NoStop}%
\bibitem [{\citenamefont {Swoboda}\ \emph {et~al.}(2002)\citenamefont
  {Swoboda}, \citenamefont {Leistam}, \citenamefont {Pigni}, \citenamefont
  {Cacaut}, \citenamefont {Kochournikov},\ and\ \citenamefont
  {Vodopyanov}}]{alice}%
  \BibitemOpen
  \bibfield  {author} {\bibinfo {author} {\bibfnamefont {D.}~\bibnamefont
  {Swoboda}}, \bibinfo {author} {\bibfnamefont {L.}~\bibnamefont {Leistam}},
  \bibinfo {author} {\bibfnamefont {L.}~\bibnamefont {Pigni}}, \bibinfo
  {author} {\bibfnamefont {D.~E.}\ \bibnamefont {Cacaut}}, \bibinfo {author}
  {\bibfnamefont {E.}~\bibnamefont {Kochournikov}}, \ and\ \bibinfo {author}
  {\bibfnamefont {A.~S.}\ \bibnamefont {Vodopyanov}},\ }\href
  {https://cds.cern.ch/record/590888} {\bibfield  {journal} {\bibinfo
  {journal} {IEEE Trans. Appl. Supercond.}\ }\textbf {\bibinfo {volume} {12}},\
  \bibinfo {pages} {432} (\bibinfo {year} {2002})}\BibitemShut {NoStop}%
\bibitem [{\citenamefont {Amapane}\ \emph {et~al.}(2005)\citenamefont
  {Amapane}, \citenamefont {Andreev}, \citenamefont {Drollinger}, \citenamefont
  {Karimaki}, \citenamefont {Klyukhin},\ and\ \citenamefont {Todorov}}]{cms1}%
  \BibitemOpen
  \bibfield  {author} {\bibinfo {author} {\bibfnamefont {N.}~\bibnamefont
  {Amapane}}, \bibinfo {author} {\bibfnamefont {V.}~\bibnamefont {Andreev}},
  \bibinfo {author} {\bibfnamefont {V.}~\bibnamefont {Drollinger}}, \bibinfo
  {author} {\bibfnamefont {V.}~\bibnamefont {Karimaki}}, \bibinfo {author}
  {\bibfnamefont {V.}~\bibnamefont {Klyukhin}}, \ and\ \bibinfo {author}
  {\bibfnamefont {T.}~\bibnamefont {Todorov}},\ }\href
  {https://cds.cern.ch/record/883293} {\  (\bibinfo {year} {2005})}\BibitemShut
  {NoStop}%
\bibitem [{\citenamefont {Klyukhin}\ \emph {et~al.}(2010)\citenamefont
  {Klyukhin}, \citenamefont {Amapane}, \citenamefont {Andreev}, \citenamefont
  {Ball}, \citenamefont {Cure}, \citenamefont {Herve}, \citenamefont {Gaddi},
  \citenamefont {Gerwig}, \citenamefont {Karimaki}, \citenamefont {Loveless},
  \citenamefont {Mulders}, \citenamefont {Popescu}, \citenamefont {Sarycheva},\
  and\ \citenamefont {Virdee}}]{cms2}%
  \BibitemOpen
  \bibfield  {author} {\bibinfo {author} {\bibfnamefont {V.~I.}\ \bibnamefont
  {Klyukhin}}, \bibinfo {author} {\bibfnamefont {N.}~\bibnamefont {Amapane}},
  \bibinfo {author} {\bibfnamefont {V.}~\bibnamefont {Andreev}}, \bibinfo
  {author} {\bibfnamefont {A.}~\bibnamefont {Ball}}, \bibinfo {author}
  {\bibfnamefont {B.}~\bibnamefont {Cure}}, \bibinfo {author} {\bibfnamefont
  {A.}~\bibnamefont {Herve}}, \bibinfo {author} {\bibfnamefont
  {A.}~\bibnamefont {Gaddi}}, \bibinfo {author} {\bibfnamefont
  {H.}~\bibnamefont {Gerwig}}, \bibinfo {author} {\bibfnamefont
  {V.}~\bibnamefont {Karimaki}}, \bibinfo {author} {\bibfnamefont
  {R.}~\bibnamefont {Loveless}}, \bibinfo {author} {\bibfnamefont
  {M.}~\bibnamefont {Mulders}}, \bibinfo {author} {\bibfnamefont
  {S.}~\bibnamefont {Popescu}}, \bibinfo {author} {\bibfnamefont {L.~I.}\
  \bibnamefont {Sarycheva}}, \ and\ \bibinfo {author} {\bibfnamefont
  {T.}~\bibnamefont {Virdee}},\ }\href {\doibase 10.1109/TASC.2010.2041200}
  {\bibfield  {journal} {\bibinfo  {journal} {IEEE Transactions on Applied
  Superconductivity}\ }\textbf {\bibinfo {volume} {20}},\ \bibinfo {pages}
  {152} (\bibinfo {year} {2010})}\BibitemShut {NoStop}%
\bibitem [{\citenamefont {Aleksa}\ \emph {et~al.}(2008)\citenamefont {Aleksa},
  \citenamefont {Bergsma}, \citenamefont {Giudici}, \citenamefont {Kehrli},
  \citenamefont {Losasso}, \citenamefont {Pons}, \citenamefont {Sandaker},
  \citenamefont {Miyagawa}, \citenamefont {Snow}, \citenamefont {Hart},\ and\
  \citenamefont {Chevalier}}]{atlas}%
  \BibitemOpen
  \bibfield  {author} {\bibinfo {author} {\bibfnamefont {M.}~\bibnamefont
  {Aleksa}}, \bibinfo {author} {\bibfnamefont {F.}~\bibnamefont {Bergsma}},
  \bibinfo {author} {\bibfnamefont {P.~A.}\ \bibnamefont {Giudici}}, \bibinfo
  {author} {\bibfnamefont {A.}~\bibnamefont {Kehrli}}, \bibinfo {author}
  {\bibfnamefont {M.}~\bibnamefont {Losasso}}, \bibinfo {author} {\bibfnamefont
  {X.}~\bibnamefont {Pons}}, \bibinfo {author} {\bibfnamefont {H.}~\bibnamefont
  {Sandaker}}, \bibinfo {author} {\bibfnamefont {P.~S.}\ \bibnamefont
  {Miyagawa}}, \bibinfo {author} {\bibfnamefont {S.~W.}\ \bibnamefont {Snow}},
  \bibinfo {author} {\bibfnamefont {J.~C.}\ \bibnamefont {Hart}}, \ and\
  \bibinfo {author} {\bibfnamefont {L.}~\bibnamefont {Chevalier}},\ }\href
  {http://stacks.iop.org/1748-0221/3/i=04/a=P04003} {\bibfield  {journal}
  {\bibinfo  {journal} {Journal of Instrumentation}\ }\textbf {\bibinfo
  {volume} {3}},\ \bibinfo {pages} {P04003} (\bibinfo {year}
  {2008})}\BibitemShut {NoStop}%
\bibitem [{\citenamefont {Jackson}(1999)}]{jackson_classical_1999}%
  \BibitemOpen
  \bibfield  {author} {\bibinfo {author} {\bibfnamefont {J.~D.}\ \bibnamefont
  {Jackson}},\ }\href {http://cdsweb.cern.ch/record/490457} {\emph {\bibinfo
  {title} {Classical Electrodynamics}}},\ \bibinfo {edition} {3rd}\ ed.\
  (\bibinfo  {publisher} {Wiley},\ \bibinfo {address} {New York, {NY}},\
  \bibinfo {year} {1999})\BibitemShut {NoStop}%
\bibitem [{\citenamefont {Overfelt}(2001)}]{overfelt}%
  \BibitemOpen
  \bibfield  {author} {\bibinfo {author} {\bibfnamefont {P.~L.}\ \bibnamefont
  {Overfelt}},\ }\href {\doibase 10.1103/PhysRevE.64.036603} {\bibfield
  {journal} {\bibinfo  {journal} {Phys. Rev. E}\ }\textbf {\bibinfo {volume}
  {64}},\ \bibinfo {pages} {036603} (\bibinfo {year} {2001})}\BibitemShut
  {NoStop}%
\bibitem [{\citenamefont {Evans}(2010)}]{evans10}%
  \BibitemOpen
  \bibfield  {author} {\bibinfo {author} {\bibfnamefont {L.~C.}\ \bibnamefont
  {Evans}},\ }\href@noop {} {\emph {\bibinfo {title} {Partial differential
  equations}}}\ (\bibinfo  {publisher} {American Mathematical Society},\
  \bibinfo {address} {Providence, R.I.},\ \bibinfo {year} {2010})\BibitemShut
  {NoStop}%
\bibitem [{\citenamefont {Waldron}(1958)}]{waldron}%
  \BibitemOpen
  \bibfield  {author} {\bibinfo {author} {\bibfnamefont {R.~A.}\ \bibnamefont
  {Waldron}},\ }\href {\doibase 10.1093/qjmam/11.4.438} {\bibfield  {journal}
  {\bibinfo  {journal} {The Quarterly Journal of Mechanics and Applied
  Mathematics}\ }\textbf {\bibinfo {volume} {11}},\ \bibinfo {pages} {438}
  (\bibinfo {year} {1958})}\BibitemShut {NoStop}%
\bibitem [{\citenamefont {MATLAB}(2018)}]{MATLAB}%
  \BibitemOpen
  \bibfield  {author} {\bibinfo {author} {\bibnamefont {MATLAB}},\ }\href@noop
  {} {\emph {\bibinfo {title} {version 9.4.0 (R2018a)}}}\ (\bibinfo
  {publisher} {The MathWorks Inc.},\ \bibinfo {address} {Natick,
  Massachusetts},\ \bibinfo {year} {2018})\BibitemShut {NoStop}%
\bibitem [{\citenamefont {Jones}\ \emph {et~al.}(01  )\citenamefont {Jones},
  \citenamefont {Oliphant}, \citenamefont {Peterson} \emph {et~al.}}]{scipy}%
  \BibitemOpen
  \bibfield  {author} {\bibinfo {author} {\bibfnamefont {E.}~\bibnamefont
  {Jones}}, \bibinfo {author} {\bibfnamefont {T.}~\bibnamefont {Oliphant}},
  \bibinfo {author} {\bibfnamefont {P.}~\bibnamefont {Peterson}},  \emph
  {et~al.},\ }\href {http://www.scipy.org/} {\enquote {\bibinfo {title}
  {{SciPy}: Open source scientific tools for {Python}},}\ } (\bibinfo {year}
  {2001--}),\ \bibinfo {note} {[Online]}\BibitemShut {NoStop}%
\bibitem [{\citenamefont {More}(1978)}]{lm_algo}%
  \BibitemOpen
  \bibfield  {author} {\bibinfo {author} {\bibfnamefont {J.}~\bibnamefont
  {More}},\ }\href {\doibase 10.1007/BFb0067700} {\bibfield  {journal}
  {\bibinfo  {journal} {Numerical Analysis}\ }\textbf {\bibinfo {volume} {630}}
  (\bibinfo {year} {1978}),\ 10.1007/BFb0067700}\BibitemShut {NoStop}%
\bibitem [{\citenamefont {Branch}\ \emph {et~al.}(1999)\citenamefont {Branch},
  \citenamefont {Coleman},\ and\ \citenamefont {Li}}]{trf_algo}%
  \BibitemOpen
  \bibfield  {author} {\bibinfo {author} {\bibfnamefont {M.}~\bibnamefont
  {Branch}}, \bibinfo {author} {\bibfnamefont {T.}~\bibnamefont {Coleman}}, \
  and\ \bibinfo {author} {\bibfnamefont {Y.}~\bibnamefont {Li}},\ }\href
  {\doibase 10.1137/S1064827595289108} {\bibfield  {journal} {\bibinfo
  {journal} {SIAM Journal on Scientific Computing}\ }\textbf {\bibinfo {volume}
  {21}},\ \bibinfo {pages} {1} (\bibinfo {year} {1999})}\BibitemShut {NoStop}%
\bibitem [{\citenamefont {Newville}\ \emph {et~al.}(2014)\citenamefont
  {Newville}, \citenamefont {Stensitzki}, \citenamefont {Allen},\ and\
  \citenamefont {Ingargiola}}]{lmfit}%
  \BibitemOpen
  \bibfield  {author} {\bibinfo {author} {\bibfnamefont {M.}~\bibnamefont
  {Newville}}, \bibinfo {author} {\bibfnamefont {T.}~\bibnamefont
  {Stensitzki}}, \bibinfo {author} {\bibfnamefont {D.~B.}\ \bibnamefont
  {Allen}}, \ and\ \bibinfo {author} {\bibfnamefont {A.}~\bibnamefont
  {Ingargiola}},\ }\href {\doibase 10.5281/zenodo.11813} {\enquote {\bibinfo
  {title} {{LMFIT: Non-Linear Least-Square Minimization and Curve-Fitting for
  Python}},}\ } (\bibinfo {year} {2014})\BibitemShut {NoStop}%
\end{thebibliography}%

\end{document}